\documentclass[twocolumn]{aa}  

\usepackage{graphicx}
\usepackage{txfonts}
\usepackage{color}
\usepackage{xcolor}
\usepackage{multicol}
\usepackage{amsmath}
\usepackage{makecell}
\usepackage{subcaption}
\usepackage{pdflscape}
\usepackage{longtable}
\usepackage{booktabs}
\usepackage{caption}
\usepackage[skip=2pt plus1pt, indent=15pt]{parskip}
\usepackage{threeparttablex}
\usepackage{tikz,caption}
\usepackage{hyperref}
\usetikzlibrary{calc}
\DeclareCaptionFormat{overlay}{\gdef\capoverlay{#1#2#3\par}}
\DeclareCaptionLabelSeparator{frcolon}{ $\,:$ }
\DeclareCaptionStyle{overlay}{format=overlay}
\usepackage{csvsimple}

\hypersetup{
    colorlinks=true,
    linkcolor=black,
    citecolor=blue,     
    urlcolor=magenta,
    linkbordercolor={red}
}
    
\begin{document}

   \title{Emergence of high-mass stars in complex fiber networks (EMERGE)}

   \subtitle{IV. Environmental dependence of the fiber widths}

   \author{A. Socci
          \inst{1}
          \and
          A. Hacar\inst{1}
          \and
          F. Bonanomi\inst{1}
          \and
          M. Tafalla\inst{2}
          \and
          S. Suri\inst{1}
          }

   \institute{1 - Institute for Astronomy (IfA), University of Vienna,
              T\"urkenschanzstrasse 17, A-1180 Vienna\\
              \email{andrea.socci@univie.ac.at} \\
              2 - Observatorio Astronómico Nacional (IGN), Alfonso XII 3, E-28014 Madrid, Spain
             }

   \date{Received; -- accepted; --}

 
  \abstract
   {Despite their variety of scales throughout the interstellar medium, filaments in nearby low-mass clouds appear to have a characteristic width of $\sim$~0.1~pc from the analysis of {\it Herschel} observations. The validity and origin of this characteristic width, however, has been a matter of intense discussions during the last decade.}
   {We made use of the EMERGE Early ALMA Survey comprising seven targets among low- (OMC-4~South, NGC~2023), intermediate- (OMC-2, OMC-3, LDN~1641N), and high-mass (OMC-1, Flame Nebula) star-forming regions in Orion, which include different physical conditions, star formation histories, mass, and density regimes. All targets were homogeneously surveyed at high-spatial resolution (4.5\arcsec or $\sim$~2000~au) in N$_2$H$^+$ (1$-$0) using a dedicated series of ALMA+IRAM-30m observations, and previous works identified a total of 152 fibers throughout this sample. Here, we aim to characterise the variation in the fiber widths under the different conditions explored by this survey.}
   {We characterised the column density and temperature radial profiles of fibers using the automatic fitting routine FilChap, and systematically quantified its main physical properties (i.e. peak column density, width, and temperature gradient).
   }
   {The Orion fibers show a departure from the isothermal condition with significant outward temperature gradients with $\nabla T_\mathrm{K} > 30$~K~pc$^{-1}$. The presence of such temperature gradients suggests a change in the equation of state for fibers.
   By fitting their radial profiles, we report a median full width at half maximum ($FWHM$) of $\sim 0.05$~pc for the Orion fibers, with a corresponding median aspect ratio of $\sim2$. Along with their median, the $FWHM$ values for individual cuts are consistently below the proposed characteristic width of 0.1~pc. More relevantly, we observe a systematic variation in these fiber $FWHM$ between different regions in our sample. We also find a direct inverse dependence of the fiber $FWHM$ on their central column density, $N_0$, above $\gtrsim 10^{22}$~cm$^{-2}$, which agrees with the expected $N_0-FWHM$ anti-correlation predicted in previous theoretical studies.}
   {Our homogeneous analysis returns the first observational evidence of an intrinsic and systematic variation in the fiber widths across different star-forming regions. While sharing comparable mass, length, and kinematic properties in all of our targets, fibers appear to adjust their $FWHM$ to their density and to the pressure in their host environment. 
   }

   \keywords{Interstellar Medium --
                Massive star formation --
                Astrochemistry
               }

   \maketitle
%
\renewcommand{\ttdefault}{pcr}
\captionsetup{labelfont=bf}

\section{Introduction}\label{sec:introduction}

Filaments are a key component of the interstellar medium (ISM). Since their first discovery \citep{barnard07}, the proximity of filaments to the young stellar objects suggested a direct connection between the filaments and the star formation process \citep[e.g.][]{hartmann02}. The far-infrared (FIR) observations from {\it Herschel} demonstrated the presence of filaments in all type of molecular clouds in our Galaxy \citep{andre10,molinari10}.

Filaments are found at all scales as part of the hierarchical structure of the ISM \citep{hacar22}.
This filamentary gas organisation extends across four orders of magnitude in scale, from the $\sim100$~pc, cloud-size giant filaments \citep[e.g.][]{jackson10,goodman14}, to the parsec-scale filament networks in nearby clouds \citep[e.g.][]{arzoumanian19} and Galactic plane surveys \citep[e.g.][]{molinari10}, down to the sub-parsec filaments within clouds \citep{hacar13}.
These latter sub-parsec filaments, known as fibers \citep{andre14}, are recognised in molecular line observations as the velocity-coherent sub-structures of parsec-scale filaments, both at low \citep[e.g.][]{tafalla15,hacar17} and high spatial resolutions \citep[e.g.][]{lee14,hacar18,dhab19}. Interferometric observations down to the $\sim1000$~au regime suggest that fibers host the dense gas prior to the formation of stars within clouds \citep{hacar24}.

While present across a wide range of scales in the ISM, \textit{Herschel} observations in nearby clouds suggest the existence of a characteristic width for filaments of $\sim0.1$~pc \citep{andre10,arzoumanian11,palme13,arzoumanian19}. An ongoing debate exists, however, about the potential variations in this previously proposed typical width of $\sim0.1$~pc. Recent results argue that the actual \textit{Herschel} widths may be distance-dependent \citep[see][]{pano22,andre22}. Beside this potential observational bias, surveys in Taurus \citep{pano14} and Orion \citep{suri19} show a variation in the filament widths up to an order of magnitude (within $0.02-0.3$~pc) in a single cloud.
Similarly, dedicated studies in regions such as OMC-3 identify different widths depending on the observational technique \citep[$\sim0.03-0.05$~pc using mid-infrared (MIR) extinction at 2\arcsec resolution;][]{juve23} and wavelength \citep[$\sim 0.06-0.08$~pc using FIR emission at 8\arcsec resolution;][]{schuller21} adopted in each case. Widths of $\sim0.03-0.05$ pc are instead systematically observed at higher spatial resolution with interferometers \citep[e.g.][]{fernandez14,hacar18,schmi21}. The additional differences in filament widths determined in massive clouds with single-dish telescopes \citep[e.g. $\sim0.3$~pc in DR21;][]{hennemann12} and interferometers \citep[e.g. $\sim0.04$~pc in NGC~6334;][]{li22} completes a broad collection of width estimates. So far, however, the large in-homogeneity of these observational studies prevents a direct comparison of these results and fuels the debate around the typical width of filaments.

The filament width is a key parameter determining the initial conditions for the star formation within these objects \citep[see][for a discussion]{pineda23}. Distinct theoretical models tried to explain the existence of a characteristic width for filaments as a result of their common formation process. 
A constant filament width could be explained in connection with the dissipative scale of turbulence \citep{padoan01,federrath16}. Filaments may also exist in quasi-equilibrium with the ambient pressure of the ISM \citep{Fiege2000}, which can confine the structure into an average 0.1~pc width \citep{fischera12}. 
These same models, however, also predict a systematic variation in the filament width with the column density at their peak and with several additional effects \citep[i.e. the external pressure, the magnetic field strength, the turbulence, and the accretion rate;][]{fischera12, heitsch13}. This dependence of the width on the column density remains as-of-yet undetected in the observations of filaments within nearby clouds \citep[see][for discussion]{arzoumanian19}. Whether a characteristic filament width is present and whether or not it may be extended to filaments in more massive and active regions within our Galaxy remains unclear. 

In this paper, we have made use of the EMERGE Early ALMA Survey, a statistically significant sample of seven low- (NGC~2023, OMC-4 South), intermediate- (OMC-2, OMC-3, LDN~1641N), and high-mass (OMC-1, Flame Nebula) star-forming regions in the  Orion A and B clouds \citep[$D\sim400$~pc; see][hereafter Paper I]{hacar24}. These targets were homogeneously surveyed at a resolution of $\theta_\mathrm{beam}\sim2000$~au (or 4.5\arcsec) with new suite of Atacama Large Millimetre Array (ALMA) plus IRAM-30m observations \citep[][Paper II]{bonanomi2024} using N$_2$H$^+$ (1$-$0) as a tracer of dense gas structure in these clouds. In a previous paper of this series, \citet{socci24} (hereafter Paper III) identified a total of 152 velocity-coherent, sonic fibers. Showing similar kinematic and mass properties, and organised in networks of increasing complexity, these fibers host the majority of protostars and appear to set the conditions for the formation of stars in these clouds. Our aim in this Paper IV of this series is to characterise the physical properties of the fiber radial profiles and their potential environmental variation.

The paper is organised as follows: we first present the theoretical description of filaments and its comparison with observational results from the literature (Sect.~\ref{sec:theorobs}); we then present the automatic fitting routine, FilChap, used to analyse the radial profiles of the Orion fibers (Sect.~\ref{subsec:oldvsnew}); third, we discuss the general properties of these radial profiles, also in connection with the algorithmic choices made to fit them (Sect.~\ref{sec:radialprofiles}, \ref{subsec:symm}); fourth, we discuss the results coming from the fitting with FilChap, starting with the fiber widths (Sect.~\ref{sec:fiberFWHM}) and corresponding aspect ratios (Sect.~\ref{subsec:arfibers}), their peak column densities (Sect.~\ref{sec:fiberN0}) and temperature gradients (Sect.~\ref{sec:fiberTKgrad}); we then explore the variation in the fiber widths throughout the survey (Sect.~\ref{subsec:systvar}), and their dependence on the column density (Sect.~\ref{subsec:coldensitydep}), also in comparison with theoretical predictions from models (Sect.~\ref{subsec:models}); we finally speculate on the dynamical state of fibers and how it may affect their width (Sect.~\ref{subsec:speculation}) and present our conclusions in Sect.~\ref{sec:conclusions}.

\begin{table*}[tbp]
    \caption{Average fiber properties across the EMERGE Early ALMA Survey.}
    \centering
    \small
    \begin{tabular*}{\linewidth}{c@{\hspace{2\tabcolsep}} | @{\hspace{2\tabcolsep}}c@{\hspace{2\tabcolsep}}c@{\hspace{2\tabcolsep}}c@{\hspace{2\tabcolsep}}c@{\hspace{2\tabcolsep}}c@{\hspace{2\tabcolsep}}c@{\hspace{2\tabcolsep}} | @{\hspace{2\tabcolsep}}c@{\hspace{2\tabcolsep}}c@{\hspace{2\tabcolsep}}c@{\hspace{2\tabcolsep}}c@{\hspace{2\tabcolsep}}c}
    \hline
    & \multicolumn{6}{c}{\textit{Paper III}} & \multicolumn{5}{c}{\textit{Paper IV}} \\
    \hline
    Source & Fibers & $M_{\rm{tot}}$ & $\sigma_\mathrm{nt}/c_\mathrm{s}$ & $L$ & \textit{m} & \textit{m}/\textit{m$_\mathrm{vir}$} & $\nabla T_{\mathrm{K}}$ & $T_\mathrm{K}^0$ & $N_0$ & $FWHM$ & $P_\mathrm{ext}/k_\mathrm{B}$ \\
     & [\#] & [M$_{\odot}$] &  & [pc] & [M$_{\odot}$ pc$^{-1}$] &  & [K pc$^{-1}$] & [K] & [$10^{22}$ cm$^{-2}$] & [pc] & [10$^6$ K cm$^{-3}$] \\
    \hline \\ [-2ex]
    OMC-1 & 30 & 209 & 0.99 & 0.094 & 26 & 0.33 & 64 & 25 & 4.9 & 0.034$^{+0.010}_{-0.007}$ & 7.7 \\ [0.7ex]
    OMC-2 & 46 & 350 & 0.66 & 0.076 & 40 & 0.90 & 76 & 19 & 6.7 & 0.043$^{+0.018}_{-0.009}$ & 3.9 \\ [0.7ex]
    OMC-3 & 24 & 257 & 0.64 & 0.085 & 50 & 1.08 & 75 & 19 & 6.4 & 0.050$^{+0.022}_{-0.010}$ & 2.6 \\ [0.7ex]
    OMC-4 South & 13 & 64 & 0.66 & 0.123 & 39 & 0.96 & 96 & 14 & 3.0 & 0.059$^{+0.019}_{-0.011}$ & 0.4 \\ [0.7ex]
    LDN~1641N & 16 & 168 & 1.09 & 0.105 & 51 & 0.97 & 150 & 17 & 4.1 & 0.063$^{+0.026}_{-0.012}$ & 1.2 \\ [0.7ex]
    NGC~2023 & 14 & 66 & 0.94 & 0.095 & 25 & 0.70 & 75 & 14 & 3.7 & 0.074$^{+0.025}_{-0.013}$ & 1.5 \\ [0.7ex]
    Flame Nebula & 9 & 94 & 0.63 & 0.111 & 27 & 0.82 & 110 & 22 & 4.2 & 0.038$^{+0.014}_{-0.013}$ & 4.0 \\ [0.5ex]
    \hline \\ [-2ex]
    Sample & 152 & 1209 & 0.74 & 0.091 & 36 & 0.78 & 80 & 19 & 5.1 & 0.05$^{+0.02}_{-0.01}$ & 3.1 \\ [0.5ex]
    \hline
    \end{tabular*}
    \label{tab:gen_prop}
\end{table*}

\section{Filament widths: Previous results}

\subsection{Theoretical predictions}\label{sec:theorobs}

The stability condition of an isothermal infinite cylinder of gas was originally studied by \citet{stodo63} and \citet{ostriker64}.
Isothermal filaments in hydrostatic equilibrium, usually referred to as Ostriker's filaments, are expected to follow a density radial profile $n(r)$ such as
\begin{equation}\label{eq:n_Ost}
  n(r)|_\mathrm{Ost} = \frac{n_0}{(1 + (r/H)^{2})^2}, \quad R_{\rm{flat}}|_\mathrm{Ost} = \sqrt{\frac{2~c_\mathrm{s}^2}{\pi~G~\mu n_0}},
\end{equation}
where $n_0$ is the gas density at the filament axis, $c_\mathrm{s}=\sqrt{\frac{k~T_\mathrm{K}}{\mu}}$ is the sound speed, $\mu$ is the mean molecular weight \citep[$=2.36$; see][]{kauff08,aspl21}, and $R_{\rm{flat}}|_\mathrm{Ost}$ is the inner flat radius uniquely determined from the balance between gravity and thermal pressure. The definition of $R_{\rm{flat}}$ in Eq.~(\ref{eq:n_Ost}) may be extended to include the effects of non-thermal motions $\sigma_\mathrm{nt}$ and magnetic fields $B$ such as $R_{\rm{flat}}'=f(c_\mathrm{s},\sigma_\mathrm{nt},B)$ \citep[see][]{hanawa1993,nakamura1993,Gehman1996}.
The integral of Eq.~(\ref{eq:n_Ost}) over all radii defines a finite critical line mass
\begin{equation}
    m_\mathrm{crit}=\int_0^{\infty} n(r)|_\mathrm{Ost}\cdot dr = \frac{2 c_\mathrm{s}^2}{G}
\end{equation}
for hydrostatic equilibrium. $m_\mathrm{crit}$ is usually employed as criterion for stability when compared with the observed line mass $m=\frac{M}{L}$ given the mass $M$ and length $L$ of a filament.

The isothermal density profile is usually generalised to the form of a Plummer-like profile:
\begin{equation}\label{eq:n_Plummer}
    n(r) = \frac{n_0}{(1 + (r/R_{\rm{flat}})^{2})^{p/2}}.
\end{equation}
Likewise, the concept of critical line mass may be extended to the virial mass $m_\mathrm{vir}=\frac{2 \sigma_\mathrm{tot}^2}{G}$, given the total gas velocity dispersion $\sigma_\mathrm{tot}^2=c_\mathrm{s}^2+\sigma_\mathrm{nt}^2$ (Paper I).

In observations, the determination of the volume density profile $n(r)$ is extremely challenging. As a consequence, the radial properties of a structure are usually determined from its (projected) column density profile $N(x)$, usually determined by integrating the volume density Plummer profile described by Eq.~(\ref{eq:n_Plummer}) along the line-of-sight \citep[see][]{arzoumanian11}:
\begin{equation}\label{eq:N_Plummer}
    N(x) = A_p\frac{n_0 R_{\rm{flat}}}{(1 + (x/R_{\rm{flat}})^2)^{(p-1)/2}}.
\end{equation}
$A_p$ is a proportionality factor depending on $p$, while the inner flat radius is linked to the observed full width at half maximum ($FWHM$) as follows:
\begin{equation}\label{eq:FWHM-Rflat}
    FWHM = 2R_{\rm{flat}}~\bigr(2^{2/(p-1)} - 1\bigr)^{1/2},
\end{equation}
which links the observed central ($x=0$) column density of the filament, $N_0$, and its corresponding volume density, $n_0$, as $R_{\rm{flat}} = \frac{N_0}{A_p n_0}$.

Equations (\ref{eq:n_Plummer}) and (\ref{eq:FWHM-Rflat}) return the Ostriker's isothermal profile for $p=4$ where $R_\mathrm{flat}=R_{\rm{flat}}|_\mathrm{Ost}$ and $FWHM\sim 1.53\times R_{\rm{flat}}|_\mathrm{Ost}$. Shallower profiles ($p<4$) are expected in other equilibrium configurations found for magnetised filaments \citep{Fiege2000}, externally pressurised filaments \cite{fischera12}, rotating filaments \citep{recchi2014}, non-isothermal filaments \citep{recchi13}, and polytropic equations of state \citep{Gehman1996,kawachi1998,toci15}, which in turn could also change the expected $FWHM$ according to Eq.~(\ref{eq:FWHM-Rflat}).

\subsection{Observations: A typical $\sim0.1$~pc width}\label{sec:theorobs}
The observational properties of the filament radial profiles are still subject to discussion both for their radial dependence, exemplified by $p$, and for their radial width, exemplified by the $FWHM$.
Studies based on \textit{Herschel} observations reported parsec-scale filaments to have a shallow column density profile with $p$ values ranging between 1.5 and 2.5 \citep{arzoumanian11,arzoumanian19}. Steeper radial profiles, close to the expected $p=4$ for an Ostriker filament (see Eq.~\ref{eq:n_Ost}), however, were identified in previous molecular observations \citep[e.g.][]{pineda2011,hacar11,monsch18,schmi21}. Similarly, \citet{hacar18} determined sharp intensity radial profiles well described by a Gaussian shape for the fibers in OMC-1 and OMC-2.

In addition to the $p$ value, those same \textit{Herschel}-based studies at low spatial resolution ($18-36''$) suggest a roughly constant width of $\sim0.1$~pc for parsec-scale filaments. This result comes from the analysis of nearby, and typically low-mass, regions part of the Herschel Gould Belt Survey \citep[HGBS;][]{arzoumanian11,andre14,arzoumanian19}, which includes the Orion B, IC5146, Aquila, Polaris, Pipe, Taurus, Musca, and Ophiuchus clouds. Nonetheless, systematic variations are identified within this HGBS sample. A factor of 2 is found between the median widths of different regions, especially when comparing Orion B and IC5146 ($FWHM\sim0.12-0.13$~pc) with the rest of the sample ($FWHM\sim0.06-0.08$~pc). While a debate exists on whether or not these widths are distance-dependant \citep{pano22,andre22}, molecular line observations in some of these regions confirm large intrinsic variations in the filament widths between $\sim0.5$~pc \citep{pano14,suri19} down to $\sim0.03$~pc depending on the tracer and resolution \citep{hacar18}.

Multiple formation and evolution scenarios could explain the relative narrow filament width distribution observed in nearby molecular clouds. 
Filaments may form from the convergence of turbulent flows, for which the dissipation scale would then become their typical width \citep{padoan01,arzoumanian11}. The same effect may be obtained for isothermal filaments in quasi-equilibrium when confined into a typical width by an external pressure \citep{Fiege2000}. Dynamically evolving filaments, either via a uniform radial velocity \citep{priest22} or via accretion \citep{heitsch13}, are expected to show a density scaling $p\lesssim4$, and a variable width along their radial profile depending on the kinematics and evolutionary stage. These model predictions suggest that the observed filament width around $\sim0.1$~pc could be explained as the average $FWHM$ from a non-isothermal and non-equilibrium condition.

\subsection{Expected dependence}\label{subsec:fwhmn0dep}

While perhaps an average value could describe the typical width of parsec-size filaments in standard low-mass clouds, theoretical calculations predict a systematic dependence of the filament $FWHM$ under different physical conditions. Following Eqs.~(\ref{eq:n_Ost}) and (\ref{eq:FWHM-Rflat}) (and its alternative formulations including turbulence and magnetic fields), the observed filament $FWHM$ should decrease for increasing central densities, $n_0$, such as $FWHM\propto T_\mathrm{K}~n_0^{-0.5}$. As a result, denser filaments are expected to show narrower widths.

The observed $FWHM$ should also correlate with the peak column density, $N_0$, for (sub-critical, i.e. $m/m_\mathrm{crit}<1$) pressure-truncated filaments in equilibrium.
This $FWHM-N_0$ dependence is controlled by the balance between the internal thermal pressure of the filament (depending on $T_{\rm{K}}$) and the external gas pressure $P_\mathrm{ext}$ at the filament surface \citep{Fiege2000}. The observed $FWHM$ predicted for these filaments reaches a maximum value $FWHM_\mathrm{max}$ at \citep[see][for a full discussion]{fischera12}
\begin{equation}
    FWHM_\mathrm{max} \sim 0.05\left( \frac{T_{\rm{K}}}{10~{\rm{K}}} \right) \left( \frac{10^{22}~\mathrm{cm}^{-2}}{N_\mathrm{0,max}} \right) \ \mathrm{pc},
    \label{eq:FWHMfischera}
\end{equation}
where the corresponding peak column density, $N_\mathrm{0,max}$, reads as
\begin{equation}
    N_\mathrm{0,max} \sim 3.5\times 10^{21}\left(\frac{P_{\rm{ext}}/k_{\rm{B}}}{2\times10^4 \mathrm{~K~cm}^{-3}} \right)^{1/2}\ \mathrm{cm}^{-2}.
    \label{eq:Nfischera}
\end{equation}
Given a fixed $P_\mathrm{ext}$, the predicted FWHM for filaments decreases around $N_\mathrm{0,max}$ for both lower column densities (i.e. $N_0<N_\mathrm{0,max}$ and low $m/m_\mathrm{crit}$) and higher column densities (i.e. $N_0>N_\mathrm{0,max}$ and large $m/m_\mathrm{crit}$).
A roughly similar behaviour is expected in the case of weakly magnetised filaments \citep{heitsch13}, while filaments confined by the ram pressure would show lower $FWHM_\mathrm{max}$ and shallower $FWHM-N_0$ variations \citep{heitsch13}. Theoretical models additionally predict filaments to occupy different positions in the $FWHM-N_0$ plane due to accretion \citep[e.g. by showing larger FWHM for larger accretion efficiencies $\epsilon$;][]{heitsch13_acc}.
Despite their differences, all these models suggest the observed $FWHM$ for filaments to decrease with column densities $N_0 > N_\mathrm{0,max}$. This predicted anti-correlation, however, has not been observed so far \citep[e.g. see discussion in][]{arzoumanian19}.

\section{Non-isothermal narrow fibers}\label{sec:radialprofiles}

\begin{figure*}[tbp]
  \centering
  \includegraphics[width=\linewidth]{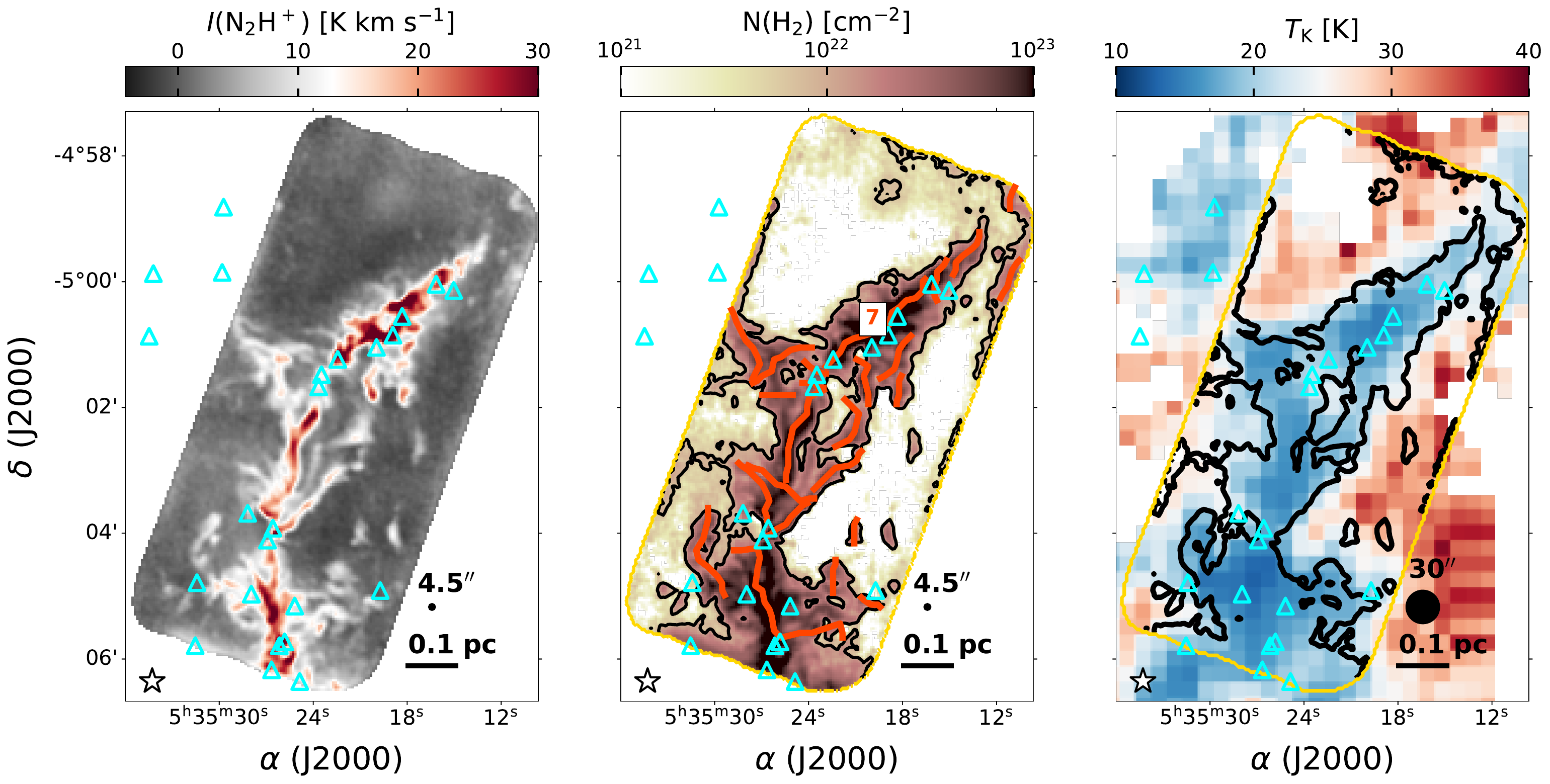}
  \caption{OMC-3 as a showcase for the results obtained from the EMERGE Early ALMA Survey (see Papers I and III for details about these maps). 
  (\textbf{Left panel}) N$_2$H$^+$ (1$-$0) integrated intensity map at a resolution of 2000~au (or $4.5$\arcsec) obtained by our ALMA+IRAM-30m observations. 
  (\textbf{Central panel}) Total column density map of H$_2$ derived at 2000~au resolution. The black contour corresponds to an intensity of N$_2$H$^+$ with S/N~=~3. Red lines define the axes of the fibers identified in Paper III where Fiber \#7 is highlighted with its ID and further discussed in Sect.~\ref{subsec:radialprofilesqual} and Fig.~\ref{fig:omc3structureprofiles}. (\textbf{Right panel}) Gas kinetic temperature $T_\mathrm{K}$ map determined at the IRAM-30m single-dish resolution of $\sim12,000$~au (or 30\arcsec). 
  In all panels, cyan triangles correspond to the protostellar objects \citep[Class 0/I;][]{megeath12,stutz13,furlan16}, while the white star corresponds to a B star \citep[gathered from the SIMBAD catalogue;][]{wenger00}.}
  \label{fig:OMC3showcase}
\end{figure*}

\begin{figure*}[tbp]
    \centering
    \includegraphics[width=0.9\linewidth]{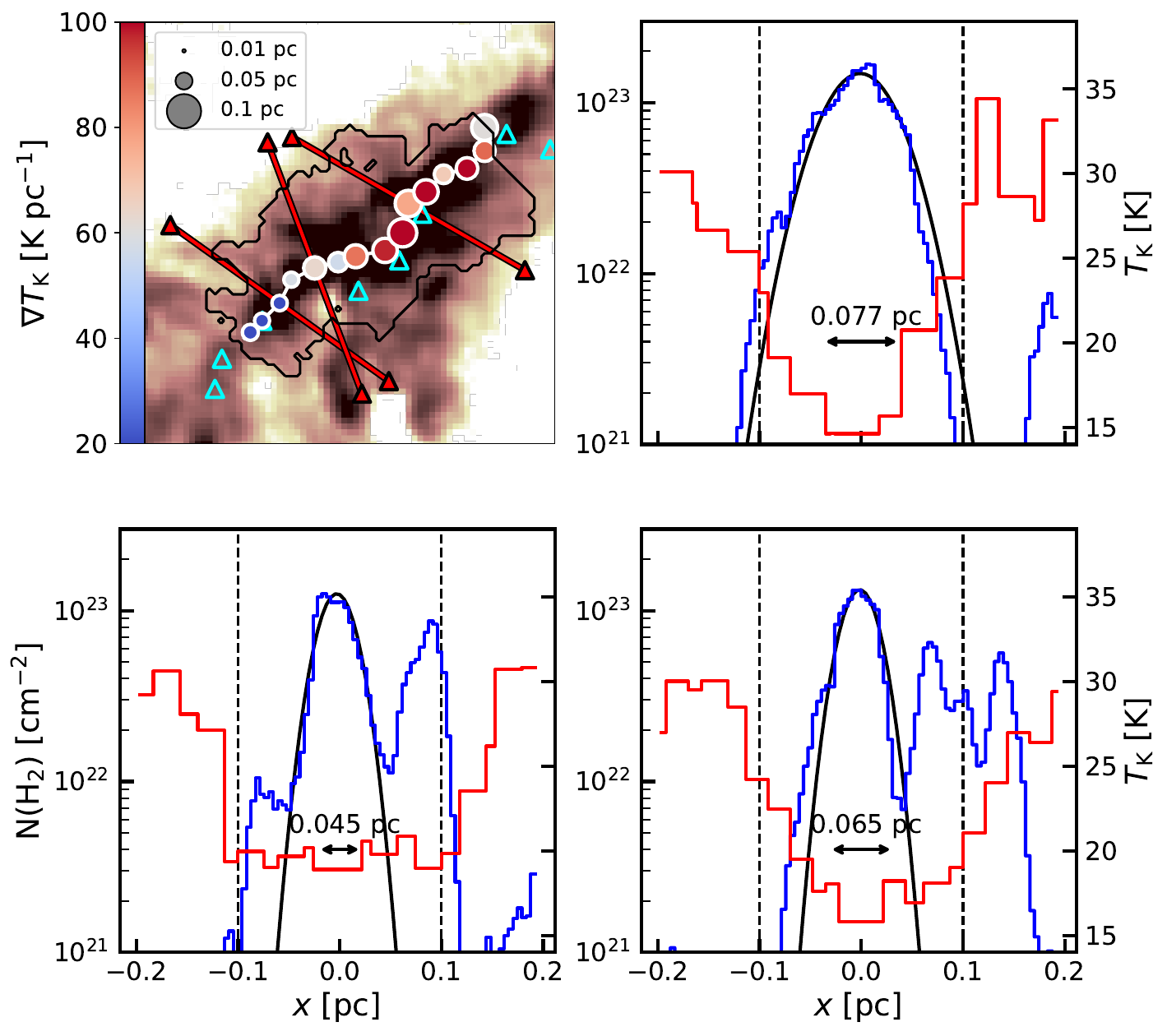}
    \caption{Radial properties measured in fiber $\#7$ in OMC-3 (see also Appendix~\ref{sec:mapsandtab}).
    \textbf{(Upper left panel)}: Zoom-in on the N(H$_2$) map presented in Fig.~\ref{fig:OMC3showcase}. All positions associated with this fiber are enclosed by a black contour. The axis of this fiber $\#7$ is indicated by a white polygon where its knots are colour-coded by the temperature gradient measured through Eq.~(\ref{eq:tgradient}) and their size corresponds to the fiber width measured at this position. Three red lines indicate the orientation and length ($\sim \pm 0.1$~pc with respect to the centre of the fiber) of the radial cuts presented in the other panels, up to the dashed lines. \textbf{(Upper right plus lower panels)} Column density (blue line) and gas kinetic temperature (red line) profiles measured perpendicular to the filament axis (centre). The best Gaussian fit to the column density profile, including its FWHM value, is indicated in each panel (solid black curve).}
    \label{fig:omc3structureprofiles}
\end{figure*}

\begin{figure*}[tbp]
    \centering
    \includegraphics[width=\linewidth]{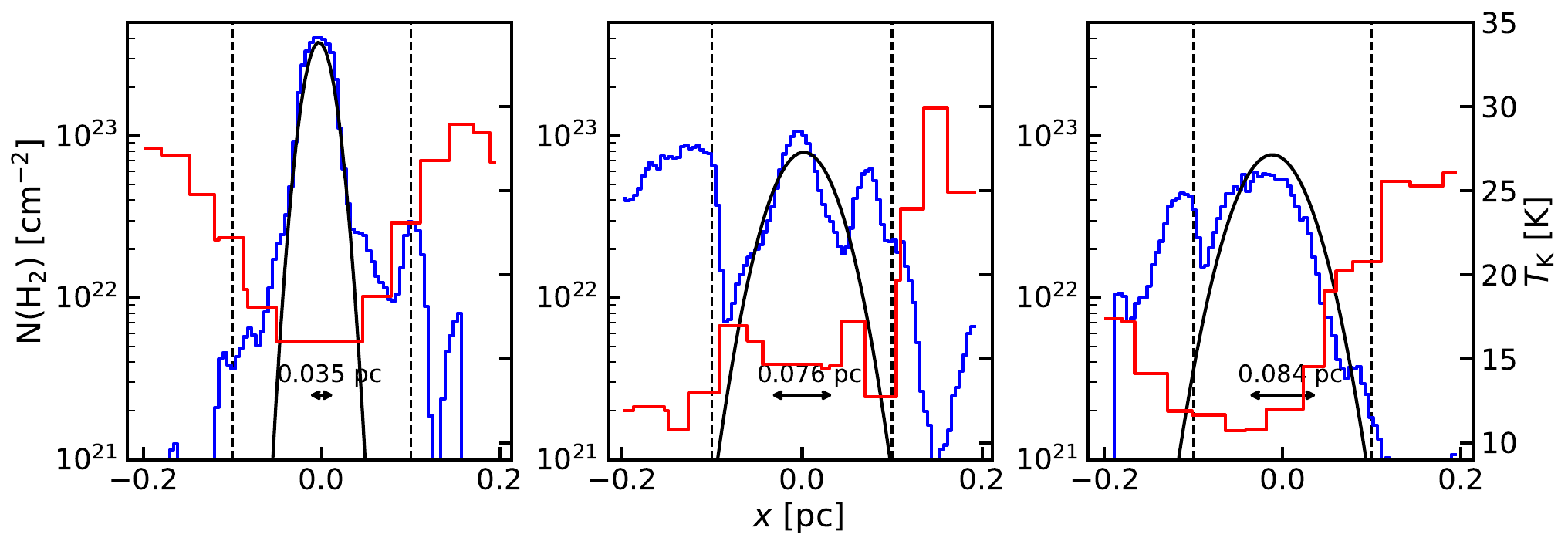}
    \caption{Radial profiles of selected cuts in three fibers belonging to the Flame Nebula (\textbf{left panel}), LDN~1641N (\textbf{central panel}), and NGC~2023 (\textbf{right panel}). Similarly to Fig.~\ref{fig:omc3structureprofiles}, we report the distance at which the temperature gradient is determined (dashed black lines) and the FWHMs determined from the best Gaussian fit (solid black line) to the density profile (blue line).}
    \label{fig:radialprofilessur}
\end{figure*}

Through the analysis of the EMERGE Early ALMA Survey we aim to statistically investigate the potential variation in the filament widths and its correlation with the peak column density $N_0$. The survey comprises seven star-forming regions in Orion surveyed at a resolution of 4.5$''$ \citep[or 2000~au at 414~pc;][]{menten07} with a new set of ALMA+IRAM-30m observations (see Paper I). We explored the dense gas organisation at sub-parsec scales under different physical and environmental conditions within these targets using the N$_2$H$^+$ (1$-$0) emission (see Paper III). The dense gas is organised in 152 velocity-coherent fibers throughout the survey, arranged in networks of different complexity depending on the physical properties of the host region. We use the analysis of the OMC-3 fibers as a showcase to highlight the most notable physical properties of these structures.

Figure~\ref{fig:OMC3showcase} (left panel) shows the total integrated N$_2$H$^+$ emission in OMC-3. The observed N$_2$H$^+$ emission with $I(\mathrm{N_2H^+})\gtrsim20$~K~km~s$^{-1}$ closely follows the parsec-scale filamentary structure already determined in previous large-scale studies \citep[e.g.][]{john99,hacar17b}. The distribution of N$_2$H$^+$ strongly correlates with the location of the Class 0/I objects identified in this region \citep{megeath12,stutz13,furlan16}, highlighting the deep connection between the dense gas traced by N$_2$H$^+$ and star formation (see also Paper I). 

The N$_2$H$^+$ map shows a high degree of complexity with several elongated sub-structures resolved at the 2000~au resolution of our ALMA+IRAM-30m observations. Compared to previous continuum studies at lower resolution which identified OMC-3 as a single filamentary structure \citep[e.g.][]{schuller21}, in Paper III we effectively identified 24 velocity-coherent, sonic fibers in OMC-3 at sub-parsec scales (red segments in Fig.~\ref{fig:OMC3showcase}, central panel). By calibrating our N$_2$H$^+$ maps against \textit{Herschel} observations in the same region, we were able to derive a high-resolution column density map for OMC-3, and our other targets (Fig.~\ref{fig:OMC3showcase}, central panel; see Paper III). Through this calibration, we revealed the N$_2$H$^+$ emission to follow the gas at highest column densities (N$(\mathrm{H_2})\gtrsim10^{22}$~cm$^{-2}$; central panel) and coldest temperatures ($T_K\lesssim$~20~K; right panel) within this cloud. Interestingly, radially increasing temperature gradients are seen in most of these structures (right panel).
Similar characteristics define the physical properties of fibers observed in other regions in this sample (see similar maps in Papers I and III). 

\subsection{Fiber radial profiles: general properties}\label{subsec:radialprofilesqual}

Before discussing quantitatively the results of our radial profile fitting, we qualitatively explored individual column density and temperature radial profiles of some individual fibers in our survey. Figure~\ref{fig:omc3structureprofiles} displays a representative sample of the radial cuts obtained during the analysis of fiber $\#7$ in OMC-3 (see also Fig.~\ref{fig:OMC3showcase}). In Fig.~\ref{fig:omc3structureprofiles} (upper left panel), we show the gas column density map associated with fiber $\#7$ (enclosed by a black contour). Individual segments indicate three selected cuts perpendicular to the main axis from which we extracted their corresponding column density (blue lines) and temperature (red lines) profiles (see upper right and lower panels). 

Figure~\ref{fig:omc3structureprofiles} highlights several important features noted during the inspection of the column density and temperature maps (see above). First, most radial profiles of fibers (blue lines) show a steep distribution in column density varying of up to two orders of magnitude within $\lesssim0.2$ pc. In addition, these radial profiles show local and global variations along the fiber axis both in number of primary and secondary peaks, and in the asymmetry of the distribution (e.g. lower left panel). Despite being undoubtedly narrower in width than previous parsec-scale filaments, individual fibers exhibit significant variations in their width along the axis (e.g. upper right panel). This intrinsic variability, already reported in the previous studies \citep{pano14,suri19}, denotes the complexity of the filamentary structures when observed at 2000~au resolution.

Second, the systematic anti-correlation between the gas column density N$\mathrm{(H_2)}$ and the gas kinetic temperature $T_\mathrm{K}$ seen in our maps is confirmed by the inspection of these individual profiles (blue and red lines, respectively; upper right and lower panels). In particular, the highest column densities within fibers usually correspond to the lowest temperatures in those same structures. This anti-correlation comes as validation of N$_2$H$^+$ being a tracer of the coldest and densest gas in our regions, enclosed by a warmer and more diffuse surrounding medium (see Paper III).

Compared to the relatively low temperatures towards the filament axis of $\sim$~20~K (in correspondence of the $N_0$ peak), the gas kinetic temperature in fibers systematically rises outwards with changes up to $\Delta T_{\mathrm{K}}>10$~K on distances of less than 0.1~pc from their axis. These temperature variations are not restricted to OMC-3, but instead they are consistently observed throughout our survey. Figure~\ref{fig:radialprofilessur} shows additional examples of radial cuts along fibers in the Flame Nebula (left panel), LDN~1641N (central panel), NGC~2023 (right panel). Similar to OMC-3, the gas kinetic temperature shows a positive radial gradient anti-correlated with the column density density distribution within single fibers. Nonetheless, more complex and asymmetric temperature profiles are also present in our sample, often influenced by the global temperature structure of the region (see Fig.~\ref{fig:radialprofilessur}, central panel). Obtained using independent observations, this local and systematic anti-correlation between N$\mathrm{(H_2)}-T_\mathrm{K}$ confirms once more the reliability of our kinematic analysis in recovering the true density structure in our targets.

\subsection{Radial fits}\label{subsec:oldvsnew}

Investigating the radial profiles in the large number of fibers included in our survey requires a systematic and reproducible analysis. Towards this end, we implemented the automatic fitting routine FilChap\footnote{ \url{https://github.com/astrosuri/filchap}} \citep{suri19} in our filament-finding algorithm to perform a full systematic characterisation of the radial profiles of the Orion fibers.
FilChap was previously employed as radial profile fitting routine in combination with different filament-finding algorithms \citep[see e.g.][]{howard19}. Its implementation in our analysis therefore required changes to the publicly available version of this software. We provide a thorough discussion on these changes in Appendix~\ref{subsec:intfilchap}, while here we only describe the main adaptations. First, we adapt FilChap to use the fiber axis identified in our kinematic analysis (see Paper III) as input for its radial analysis.
Second, we simultaneously feed FilChap with both the column density and the temperature maps (e.g. Fig.~\ref{fig:OMC3showcase}). The column density map is sampled to determine the average density profile along the axis of each structure, which is then fitted with different orthogonal cuts; later on, the temperature map is sampled along the same radial cuts without performing any average to give an estimate of the temperature gradient (see Sect.~\ref{sec:fiberTKgrad}). And third, we avoid any baseline subtraction, usually applied by FilChap, since we removed the column density floor term when converting the $I(\mathrm{N_2H^+})$ emission into the total N$(\mathrm{H_2})$ column density in our maps (see Paper III for a discussion). 

FilChap considers a series of radial cuts across the main axis of each fiber in our survey. For each section of this axis, FilChap performs a sampling of the input column density and temperature maps using a perpendicular cut. The first and second derivatives of the (average) resulting profile are then evaluated point-by-point to determine local extremes. These are labelled as either peaks or shoulders depending on a user-defined limit of significance. The maximum enclosed within two minima, and closest to the axis segment, is considered as the filament ridge and is fitted with a Plummer-like (i.e. following Eq.~\ref{eq:N_Plummer} with either $p=2$ or $p=4$ values) and a Gaussian distribution (see discussion below). FilChap then returns the full radial profile distribution (both in N(H$_2$) and $T_\mathrm{K}$) and the main radial profile features identified in the sampling (e.g. number of peaks, shoulders), along with the best fit results (e.g. Gaussian parameters) and quality assessments.

The additional parameters provided by FilChap allowed us to evaluate the best fitting approach for the Orion fibers. A full discussion on the tests and the related statistics is given in Appendix~\ref{subsec:symm}, while here we only report the main conclusions. We first inspected the skewness and excess kurtosis of the cuts, which suggest symmetric profiles, on average, with a sharp drop in column density from their central peak. Both features were qualitatively noted from the column density maps (Fig.~\ref{fig:OMC3showcase}), and profiles (Fig.~\ref{fig:omc3structureprofiles}), and denote the intrinsically complex gas organisation resolved by ALMA. 
We then investigated the number of peaks and shoulders identified by FilChap throughout our sample of profiles. The majority ($\sim90\%$) of the selected cuts have only one significant maximum within the radially sampled distance, suggesting a single-peaked function as best fitting profile. We finally explored the different fitting functions provided by FilChap (i.e. Gaussian and Plummer profiles, see Sect.~\ref{sec:theorobs}). The fitted radial profiles widths (as $FWHM$) agree well with median values of $\sim0.04-0.05$~pc and show similar distributions for the functions within $\sim0.01-0.3$~pc. The Gaussian profile, however, is our function of choice for the following reasons: first, it provides the least number of unresolved widths (i.e. $FWHM<0.027$~pc, $3\times\theta_\mathrm{beam}$ at 414~pc); second, it returns the lowest dispersion and median value for the reduced $\chi^2$; third, it has the lowest degeneracy given the lower number of free parameters (3) compared to the Plummer profile (4), which compelled us to enforce the $p$-value (see Sect.~\ref{sec:theorobs}). We shall therefore only discuss the Gaussian fit results hereafter.

Applied to the total 152 fibers identified in our EMERGE sample, FilChap retrieved a total of 1052 radial cuts along their axes, each including measurements of both the total gas column density (Sects.~\ref{sec:fiberFWHM}-\ref{sec:fiberN0}) and gas kinetic temperature (Sect.~\ref{sec:fiberTKgrad}) profiles. FilChap was able to provide a convergent fit for 991 of these cuts, which correspond to 150 objects with at least one non-empty cut.
For each of these objects, FilChap allows us to investigate their FWHM (Sect.~\ref{sec:fiberFWHM}), aspect ratio (Sect.~\ref{subsec:arfibers}), peak column density (Sect.~\ref{sec:fiberN0}), and temperature gradient (Sect.~\ref{sec:fiberTKgrad}).

\subsection{Fiber FWHMs below 0.1~pc}\label{sec:fiberFWHM}

\begin{figure*}[ht]
\centering
\includegraphics[width=\linewidth]{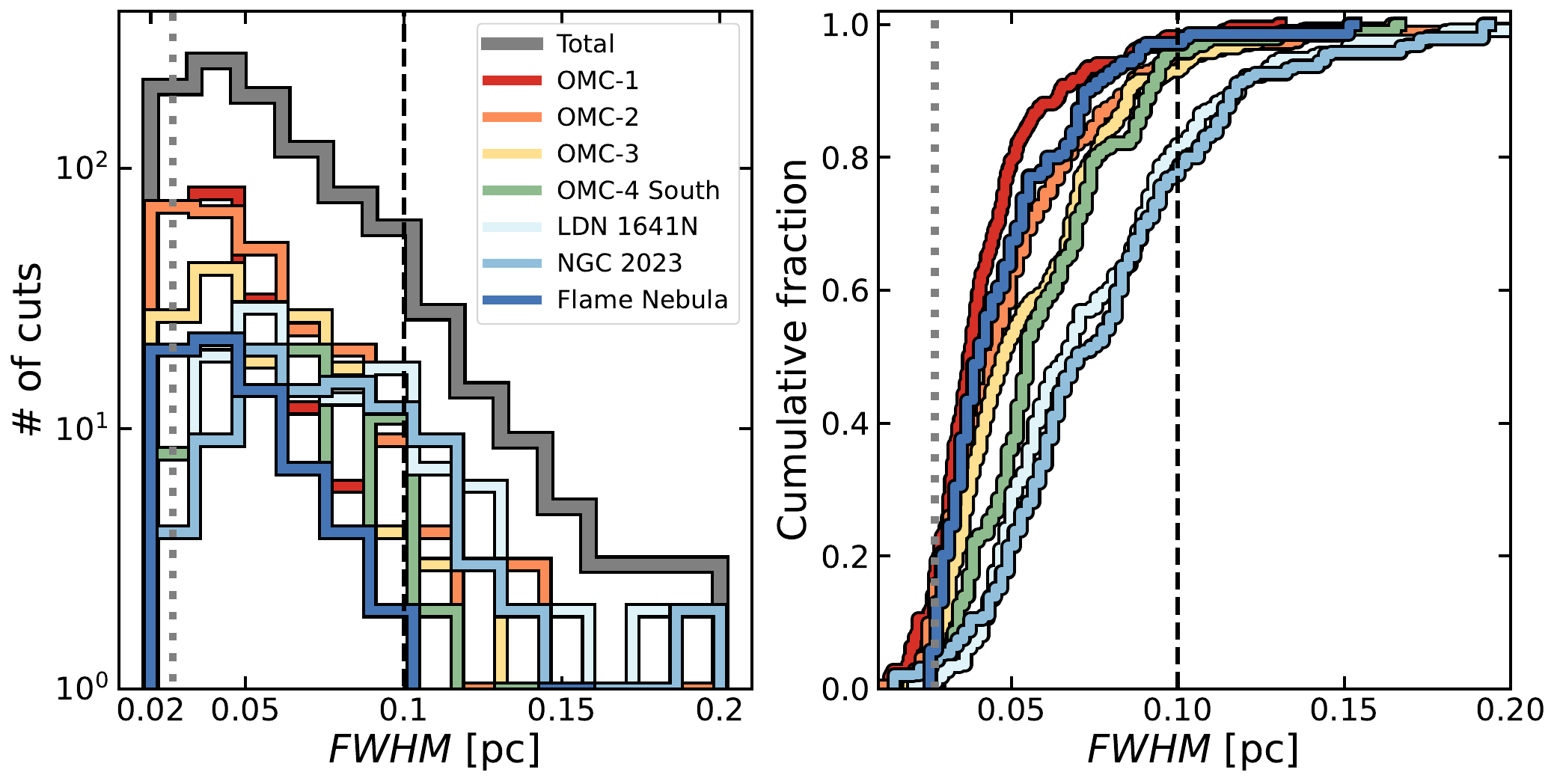}
\caption{Distribution for the FWHM derived across all the cuts in each fiber radial profile. \textbf{(Left panel)} Histogram of the fiber FWHM divided per region. \textbf{(Right panel)} Cumulative distribution of the same FWHM per fiber cut.
The dashed black line in the left panel corresponds to the characteristic FWHM measured in parsec-scale filaments in the Solar neighbourhood  \citep[e.g.][]{arzoumanian11} while a dotted grey line indicates the minimum size ($3\times \theta_{beam}$) for our fibers to be considered as resolved by our observations.}
\label{fig:cumulativerad}
\end{figure*}

We display the $FWHM$ values measured in our Gaussian fits in Fig.~\ref{fig:cumulativerad}. Rather than averaged per fiber (see also Sect.~\ref{subsec:systvar}), we first opted to display all individual fit values for the total 991 radial cuts across our fibers. 
The reason behind this choice is twofold: first, all fibers show large internal variations with the $FWHM$ varying significantly along different cuts within the same structure (see also Appendix \ref{sec:mapsandtab}); second, this approach takes advantage of our large statistical sample to account for the full intrinsic scatter of these measurements \citep[see][for a discussion]{panopoulou2017}. Given the different number of fibers per region (see Table \ref{tab:gen_prop}), the corresponding number of radial cuts is however uneven.

As is seen in Fig.~\ref{fig:cumulativerad} (left panel), the widths of the Orion fibers span more than an order of magnitude in $FWHM$, between $\sim0.01$~pc ($\sim \theta_\mathrm{beam}$) and $\sim0.2$~pc. 
The corresponding median width is $FWHM=0.05^{+0.02}_{-0.01}$~pc, which agrees well with those determined in other molecular line studies at comparable resolution \citep[e.g.][]{hacar18,schmi21,li22}. On the other hand, this $FWHM$ is a factor of 2 lower than the previously proposed characteristic $\sim$~0.1~pc width for parsec-scale filaments \citep[see vertical dashed line;][]{arzoumanian11}. Within our sample, only 4\% of our cuts show a $FWHM$ above 0.1~pc, a fraction which varies with the region: $\sim20\%$ of the cuts in NGC~2023 reach widths above 0.1~pc, while $<10\%$ are found above this value in OMC-1 and the Flame Nebula (see also right panel) where the fibers show instead a typical width of $\sim0.04$~pc. With only 5\% of the cuts apparently unresolved at our ALMA resolution (i.e. $FWHM<3\times \theta_\mathrm{beam}\sim0.027$~pc), our analysis is robust and demonstrates a clear differentiation between the width of filaments at parsec-scales compared to the width of fibers at sub-parsec scales.

Among the regions in the survey, the results in OMC-3 deserve additional remarks. The $FWHM$ of the filaments at different scales in this region was recursively explored at high resolution by multiple independent studies. \citet{schuller21} measured shallow profiles ($p\sim 2$) with a $FWHM$ of $\sim0.06$~pc along the main spine of OMC-3 from the analysis of the dust continuum emission (at 8\arcsec resolution). \citet{juve23} reported instead steep profiles ($p\sim 3$) in OMC-3 with a median $FWHM$ of $\sim0.05$~pc using MIR dust extinction (at 2\arcsec resolution). Our line emission measurements in this region (at 4.5\arcsec resolution) report a median $FWHM$ of 0.05~pc (see Table \ref{tab:gen_prop}), which is compatible with these previous results. Despite the different tracers (continuum vs lines; see further discussion later), the close agreement between these different observational results suggests these widths to be characteristic of the region and confirms the ability of our high-resolution maps to sample the true column density distribution in our fibers.

\begin{figure}[ht]
\centering
\includegraphics[width=\linewidth]{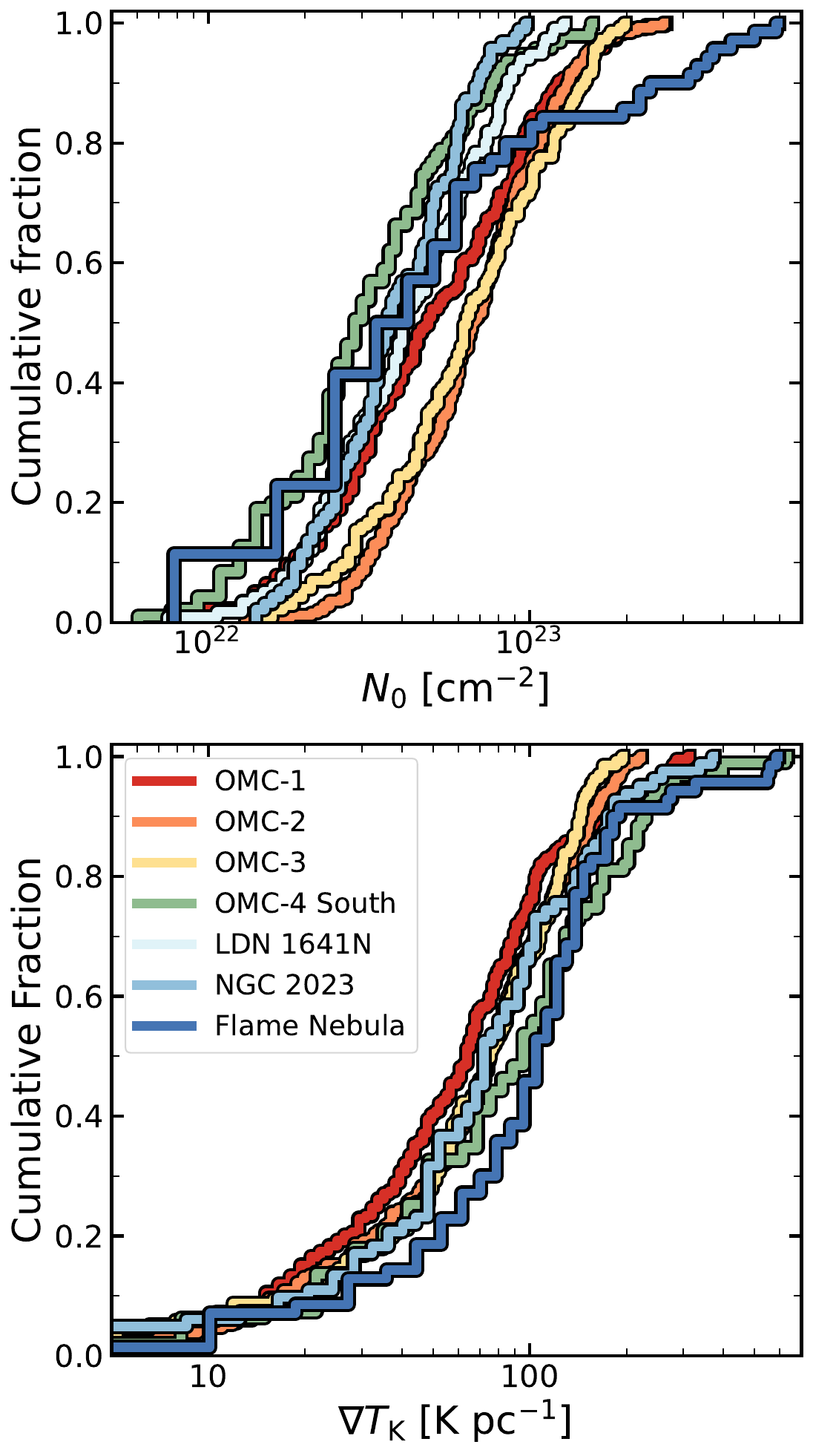}
\caption{Cumulative distributions of the (\textbf{Upper panel}) peak column densities (N$_0$) and (\textbf{Lower panel}) temperature gradients ($\nabla T_\mathrm{K}$; see Eq.~\ref{eq:tgradient}) derived for all the radial cuts in our sample.}
\label{fig:cumulativeNT}
\end{figure}

\subsection{Aspect ratios}\label{subsec:arfibers}

Fibers previously identified in OMC-1 and OMC-2 showed aspect ratios within $\sim3-7$ \citep{hacar18}. Compared to these estimates, our analysis recovers fibers with lower aspect ratios (defined as $AR = L/FWHM$ using the widths determined with FilChap) in these same two regions, as well as in the additional five clouds composing this survey ($\langle AR\rangle \sim 2$ throughout the survey; see Paper III for a discussion). The reason behind the different $AR$ values is the choice of column density thresholds used in our new fiber identification scheme. Needed to homogeneously recover the wide range of properties sampled in the EMERGE Early ALMA Survey (Paper I), the larger number and wider range of column density thresholds used in the re-analysis of these data (6 thresholds between 20-120~$\mathrm{A_V}$; see Paper III) leads to the fragmentation (and thus the decrease in their $L$ and $AR$) of several of the OMC-1/-2 fibers previously identified as monolithic structures by \citet{hacar18} (using instead three thresholds between 26-66~$\mathrm{A_V}$). 

Although arbitrary, we stress that our algorithmic choices, column density thresholds included, do not alter our the conclusions obtained from our $FWHM$ measurements. This remains true even if we remove those fibers with $AR<3$, aspect ratio typically used to define filaments \citep[e.g.][]{arzoumanian19}: although significantly reduced in statistics (only 40 available), the Orion fibers still show widths of $\sim0.03-0.04$~pc on average.

\subsection{Peak column densities above 100~A$_\mathrm{V}$}\label{sec:fiberN0}

We explore now the peak column densities of each cut as the other main output of FilChap. Figure~\ref{fig:cumulativeNT} (upper panel) displays the cumulative distributions of $N_0$, the amplitude of the Gaussian fitted to our individual radial cuts. $N_0$ spans two orders of magnitude, from $\sim10^{22}$~cm$^{-2}$ ($\sim10$~A$_\mathrm{V}$) up to $\sim5\times10^{23}$~cm$^{-2}$ ($\sim500$~A$_\mathrm{V}$). With the exception of the Flame Nebula, all the regions in our survey show similar cumulative distributions, smoothly increasing in the parameter space. OMC-1, OMC-2, and OMC-3, however, display distributions slightly skewed towards higher $N_0$ values, on average. This behaviour is expected from the large number of fields showing column densities N$(\mathrm{H_2})\gtrsim2\times10^{22}$~cm$^{-2}$ in these three regions, which further influences their total dense gas mass ($M_\mathrm{tot}\gtrsim200$~M$_{\odot}$; see Table~\ref{tab:gen_prop}).

Not surprisingly, our sub-parsec fibers show systematically higher column densities than those reported for filaments in nearby clouds. Our EMERGE Early ALMA Survey shows a median value of $N_0\sim5\times10^{22}$~cm$^{-2}$, or an equivalent extinction peak of $\sim50$~A$_\mathrm{V}$. These median values are an order of magnitude higher than those reported in the analysis of the filaments part of the HGBS \citep[$\sim5\times10^{21}$~cm$^{-2}$;][]{arzoumanian19}. This difference is expected due to the combination of several selection effects. Compared to the HGBS, our ALMA survey focuses on the Orion A and B clouds, the two most massive clouds in the Solar neighbourhood. Within these clouds, our observations further target those regions with bright N$_2$H$^+$ emission, which are associated to large factions of dense gas (see Paper I). Finally, our ALMA+IRAM-30m maps achieve a resolution that is better by factor of $\sim5$ (4.5\arcsec vs $18-36''$), allowing to resolve the dense gas substructure within the parsec-scale filaments, which is otherwise smeared out. The differences reported in $FWHM$ (see Sect.~\ref{sec:fiberFWHM}) and $N_0$ (this section) highlight the importance of these high-resolution observations in order to investigate the complex gas organisation in regions such as Orion.

\subsection{Non-isothermal fibers}\label{sec:fiberTKgrad}

Usually overlooked, the thermal profile of filaments has a significant impact on their physical evolution.
Most studies compared the observed radial distributions with those of isothermal filaments (i.e. the Ostriker filament; Eq.~\ref{eq:n_Ost}). Compared to this classical isothermal configuration, different theoretical works demonstrate how radial temperature gradients can change the pressure balance within filaments. An imbalance in the pressure further alters the radial distribution of gas within the filament at equilibrium, and the corresponding critical line mass \citep[e.g.][]{recchi13}. Additional variations in this equilibrium are expected in the case of polytropic, and non-isothermal, equations of state \citep{Gehman1996,kawachi1998,toci15}. 
Observational works report temperature variations of a few Kelvin on radial distances of $\sim0.1-0.5$~pc in only a selected number of targets \citep[e.g.][]{stepnik03,palme13,bonne20}. No systematic study of these gas temperature changes within filaments, however, has been performed so far. Our gas kinetic temperature maps (see Fig.~\ref{fig:OMC3showcase}, right panel) thus offer a unique opportunity to characterise the thermal structure around and within fibers in a large statistical sample. 

The temperature profiles of the Orion fibers show a large variation in morphology, especially when looking at individual cuts (see Figs.~\ref{fig:omc3structureprofiles},~\ref{fig:radialprofilessur}). The combination of complex profiles, which are often asymmetric, and the low resolution of the observations prevents us from performing a fit of these temperature radial distributions. As a first-order characterisation of these profiles, we instead estimated the temperature gradient in each radial cut as follows: we compared the temperature difference at the column density peak, $T_\mathrm{K}^0$, with the average temperature outside our fibers $T_\mathrm{bg}$ at a radial distance, $\pm~\Delta x$, as
\begin{equation}
    \nabla T_{\rm{K}} = \frac{T_{\rm{bg}} - T_\mathrm{K}^0}{\Delta x},
    \label{eq:tgradient}
\end{equation}
where $\nabla T_{\rm{K}}$ measures the gas temperature gradient in units of K~pc$^{-1}$.
Given the typical widths of the Orion fibers (see also Sect.~\ref{sec:fiberFWHM}), we chose a radial distance of $\pm0.1$~pc from the column density peak as representative of the environment outside the fiber. This choice allowed us to study the temperature variations along at least three beams of the IRAM-30m maps (30$''$, or $\sim0.06$~pc). Figure~\ref{fig:cumulativeNT} (lower panel) shows the temperature gradients per cut in our survey sampled in logarithmic intervals.

The whole survey shows a distribution of temperature gradients ranging two orders of magnitude within $\sim5-500$~K~pc$^{-1}$. While differences exist between distributions (e.g. Flame Nebula vs OMC-4 South), all regions have the majority of cuts with temperature gradients $> 30$~K~pc$^{-1}$. As a consequence, the median gradients for the seven region in the survey are $\nabla T_{\rm{K}}\gtrsim80$~K~pc$^{-1}$ (see Table \ref{tab:gen_prop}). This value is an order of magnitude larger than the dust temperature gradients of $\nabla T_{\rm{dust}}\sim5$~K~pc$^{-1}$ observed in filaments such as B213/L1495 \citep{palme13} and Musca \citep{bonne20}. Gas kinetic temperatures higher compared to the dust temperatures (i.e. a higher thermal pressure component) are expected at low densities due to the combination of inefficient gas cooling and weak dust-to-gas thermal coupling \citep{goldsmith2001}. 
Compared to these previous observations of filaments in quiescent regions, such as Taurus and Musca, the temperature gradients reported in our survey suggest that fibers are non-isothermal structures in regions with intense stellar activity, such as Orion. 
Non-isothermality should be therefore considered when interpreting the formation, evolution, and initial conditions for star formation of filaments in the ISM.

\section{Environmental variation in the fiber widths}\label{subsec:systvar} 

As is discussed in Sect.~\ref{sec:fiberFWHM}, the fiber widths in our survey show a median $FWHM$ of 0.05$^{+0.02}_{-0.01}$~pc (using the interquartile range, IQR, as uncertainty). 
This median value is marginally compatible, within the errors, with the $FWHM$ reported for parsec-scale filaments in the nearby clouds part of the HGBS \citep[$0.09^{+0.04}_{-0.03}$~pc;][]{arzoumanian19}. An inspection of Fig.~\ref{fig:cumulativerad}, however, reveals our survey samples lower widths, on average, with the vast majority of the Orion fibers showing $FWHM\lesssim0.1$~pc. These deviations from a typical fiber width of 0.1~pc were already reported in OMC-1 and OMC-2 alone by \citet{hacar18} and are now extended to all seven regions in the EMERGE Early ALMA Survey. 

We additionally observe a systematic variation in the fiber widths from region to region in our survey.
This effect is clearly visible in Fig.~\ref{fig:cumulativerad} (right panel) as a progressive increase in the skewness of the cumulative distributions for the $FWHM$ in the different targets composing our survey. Regions such as OMC-1, OMC-2, and the Flame Nebula have the narrowest individual cuts, with a median value of $\sim0.03-0.04$~pc. On the other hand, NGC~2023 and LDN~1641N show larger values up to $0.06-0.07$~pc (see Table \ref{tab:gen_prop}). 
The density-selective nature of N$_2$H$^+$ as gas tracer has been invoked to explain the differences between the widths measured in millimetre line observations compared to those observed in FIR continuum \citep{Priestley2023}. Our EMERGE Early ALMA survey was designed to minimise this bias by providing a sample of targets located at the same distance and all systematically observed with a uniform resolution (see Paper I). Relative differences within our survey cannot be attributed to observational biases as they would homogeneously affect all of our targets. Instead, the progressive variation in width observed in Fig.~\ref{fig:cumulativerad} must correspond to a change in the (average) physical properties and environmental conditions of the fibers in the regions surveyed. 

Our results present evidence of a systematic variation in the observed fiber $FWHM$ as function of the environment. Both local (Sect.~\ref{sec:fiberFWHM}) and region-wide variations (see above) critically challenge the existence of a unique and characteristic width for filaments and fibers \citep{andre14}. These variations suggest instead a scenario in which filaments present a wide range of widths in connection to the properties of their host region. In the following sections, we shall explore the physical origin of these different $FHWM$ values within our fibers.

\subsection{$FWHM-N_0$ anti-correlation}\label{subsec:coldensitydep}

\begin{figure*}
\centering
\includegraphics[width=0.8\linewidth]{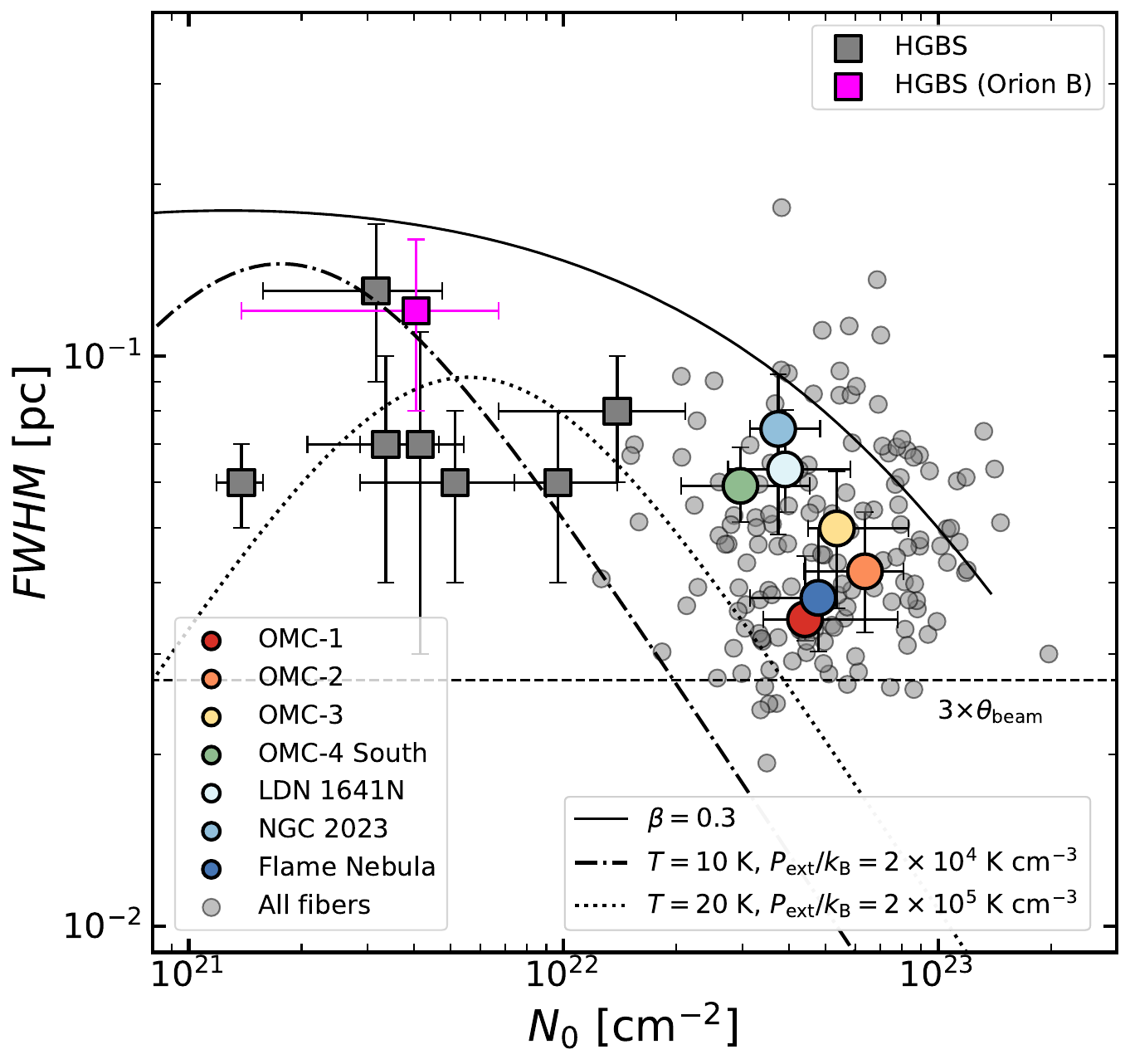}
\caption{Comparison between the peak column densities ($N_0$) and widths ($FWHM$) of the fibers in the EMERGE Early ALMA Survey (circles, this work), the \textit{Herschel} filaments in the HGBS \citep[squares,][]{arzoumanian19}. The large circles represent the median widths and densities per region in our survey. The dashed black line indicates the limit for unresolved fibers in our study ($3\times\theta_\mathrm{beam}$, or 0.027~pc). Different curves denote representative examples of the $FWHM-N_0$ dependence expected in pressure-truncated filaments with $(T_\mathrm{K}, P_\mathrm{ext}/k_\mathrm{B})=(10~\mathrm{K}, 2\times10^4$~K~cm$^{-3}$; dash-dotted line) and $(T_\mathrm{K}, P_\mathrm{ext}/k_\mathrm{B})=(20~\mathrm{K}, 2\times10^5$~K~cm$^{-3}$; dotted line), respectively \citep{fischera12}, as well as those for weakly magnetised filaments ($\beta=0.3$) \citep{heitsch13}.}
\label{fig:filcompar}
\end{figure*}

There is no typical width for fibers in our survey, but instead a systematic shift of its value from region to region. This variation suggests the influence of external environmental conditions and/or a change in the physical properties of fibers, the column density being the best candidate among the latter. A column density dependence of the filament and fiber width, although expected from theoretical models \citep[e.g.][]{fischera12,heitsch13}, has never been determined in previous surveys of nearby clouds \citep[e.g.][]{suri19,arzoumanian19}. We shall now explore if such a correlation exists and if the theoretical predictions can reproduce it.

Figure~\ref{fig:filcompar} shows the $FWHM$ and $N_0$, obtained as median values of those in individual cuts, for each fiber in our survey (grey circles). The use of these median values reduces our dynamic range in column density to an order of magnitude, between $\sim10^{22}$~cm$^{-2}$ and $\sim10^{23}$~cm$^{-2}$. Within this range, the widths also vary by almost an order of magnitude, with the majority of fibers showing values between $\sim0.02$ (barely resolved at our resolution) and $\sim0.1$~pc. Despite the scatter, the fibers are not randomly distributed in the $FWHM-N_0$ parameter space. To later compare with the HGBS from \citet{arzoumanian19}, we also computed the average values per region (colour-coded circles) with the corresponding errors. These averages show a clear anti-correlation of the width with the column density. The most massive regions ($M_\mathrm{tot}\gtrsim250$~M$_{\odot}$) and with a significant number of protostars ($\mathrm{P}\gtrsim20$), such as OMC-2 and OMC-3, occupy the bottom right corner. On the other hand, regions with lower masses ($M_\mathrm{tot}\lesssim150$~M$_{\odot}$) and a lower protostellar content ($\mathrm{P}\lesssim15$), such as OMC-4 South, NGC~2023, and LDN~1641N, are located in the top left corner. More evolved regions, such as OMC-1 and the Flame Nebula, show peak column densities similar to those of OMC-2 and OMC-3, but with narrower widths on average (see Paper I/III for the general properties of the regions). The feedback from the OB stars hosted within these two high-mass regions possibly promotes the narrowing of the fibers through a combination of higher external pressures, the dense gas ablation or dispersal, or the chemical destruction of N$_2$H$^+$ \citep[see e.g.][]{tafalla23}. Despite these possible and various influences throughout our targets, the column density appears to be the prime factor determining the fiber width.

In Paper III, we saw how the dense gas probed by our observations closely follow the total column density seen by \textit{Herschel}. Although at different resolutions (4.5$''$ and 18$''$, respectively), we try to compare the results obtained from our study and the HGBS \citep[see Table 3 in][]{arzoumanian19} by adding the filament properties they derived per region to Fig.~\ref{fig:filcompar} (grey squares). The combined distribution of the two surveys ranges two orders of magnitude in $N_0$, uniformly sampling the parameter space within $\sim10^{21}-10^{23}$~cm$^{-2}$. The combined widths of the two surveys show a continuous and constant value around $\sim0.06-0.07$~pc up to $N_0\sim3\times10^{22}$~cm$^{-2}$, with the only exceptions being Orion B and IC5146 with $FWHM\sim0.12-0.13$~pc. It is worth noting that these two regions are the farthest in the HGBS sample, and therefore possibly retain higher widths because of a distance bias \citep{pano22}. The combined distribution shows instead a systematic decrease in width as $N_0$ grows above $\gtrsim4\times10^{22}$~cm$^{-2}$. In this column density regime, OMC-1/-2/-3 and the Flame Nebula show $FWHM$ values below $\sim0.07$~pc, down to $\sim0.034$~pc, progressively decreasing for increasing $N_0$.

\subsection{Comparison with the theoretical models}\label{subsec:models}

In Figure~\ref{fig:filcompar} we display, along with the regions surveyed, the $FWHM-N_0$ relations predicted by a few representative models for pressure-confined and accretion-truncated filaments (see Sect.~\ref{subsec:fwhmn0dep}). Being $N_0$ the peak column density of H$_2$, we corrected by a factor of 2 the models from \citet{fischera12}, which were originally determined for the density of H nucleons. A first model (dash-dotted line) describes the average conditions for the cold ISM ($T_{\rm{K}}=10~\mathrm{K}$, $P_\mathrm{ext}/k_\mathrm{B}=2\times10^4$~K~cm$^{-3}$), representative for the HGBS filaments \citep{fischera12}.
A second model (dotted line) describes conditions closer, both in temperature and external pressure, to those expected for our sample ($T_{\rm{K}}=20~\mathrm{K}$, $P_\mathrm{ext}/k_\mathrm{B}=2\times10^5$~K~cm$^{-3}$; see Sect.~\ref{subsec:speculation}).
Finally, the third model (solid line) illustrates the $FWHM$ variations in weakly magnetised filaments \citep[with plasma factor $\beta=0.3$;][]{heitsch13}. Interestingly, the HGBS filaments exhibit peak column densities with $N_0\lesssim 10^{22}$~cm$^{-2}$, which corresponds to the flat regime in the above $FWHM-N_0$ correlations. When compared to the model predictions, the limited dynamic range in column density of the HGBS prevents the detection of further variations in the $FWHM$ for $N_0$ above $\gtrsim10^{22}$~cm$^{-2}$. The inclusion of our measurements for this column density regime highlights the drop in filament width for increasing column density, as was expected from the theoretical predictions. This result remains robust even considering only those fibers with $AR > 3$, whose scatter in the $FWHM-N_0$ parameter space is reproduced by the models. While none of these models can fully constrain the observations, the combined distribution of the two surveys achieves a dynamic range in column density broad enough to reproduce the predicted $FWHM$ variation with $N_0$.

Despite showing a systematic $FWHM-N_0$ anti-correlation for filaments at high column densities, the interpretation of these results is complex. The above theoretical models can only describe sub-critical filaments ($m/m_\mathrm{crit}<1$), usually cast as isothermal infinitely long cylinders. The fibers identified in the EMERGE Early ALMA Survey appear instead as non-isothermal, prolate spheroids, and super-virial in a few occurrences. In addition, projection effects can introduce significant uncertainties to the expected $N_0$ values \citep{fischera12}, challenging the applicability of these models to the Orion fibers. Nonetheless, Fig.~\ref{fig:filcompar} represents a first observational evidence that these changes fiber widths could be related to their host environments: massive regions rich in (proto-)stars (OMC-1/2/3, Flame Nebula) appear to populate higher $N_0$ and lower $FWHM$ regimes, on average, while regions with lower masses and a lower number of (proto-)stars (OMC-4 South, NGC~2023, LDN~1641N) populate the opposite regimes instead. This stratification could be explained by the combined effects of increasing external pressures (both thermal and non-thermal), of increasing self-gravity towards high-density regimes, and of different degrees of thermal support promoted by the non-isothermality of the Orion fibers. All of these influences likely induce the global scatter seen in Fig.~\ref{fig:filcompar} around the $FWHM-N_0$ distribution expected from simplified models (see dash-dotted line in this figure). Additional observations are needed to constrain key parameters assumed in these models, such as the external gas density and the accretion rates onto fibers \citep[see][for a discussion]{heitsch13}.

\subsection{Are fibers pressured confined?}\label{subsec:speculation}

\begin{figure}[ht]
\centering
\includegraphics[width=\linewidth]{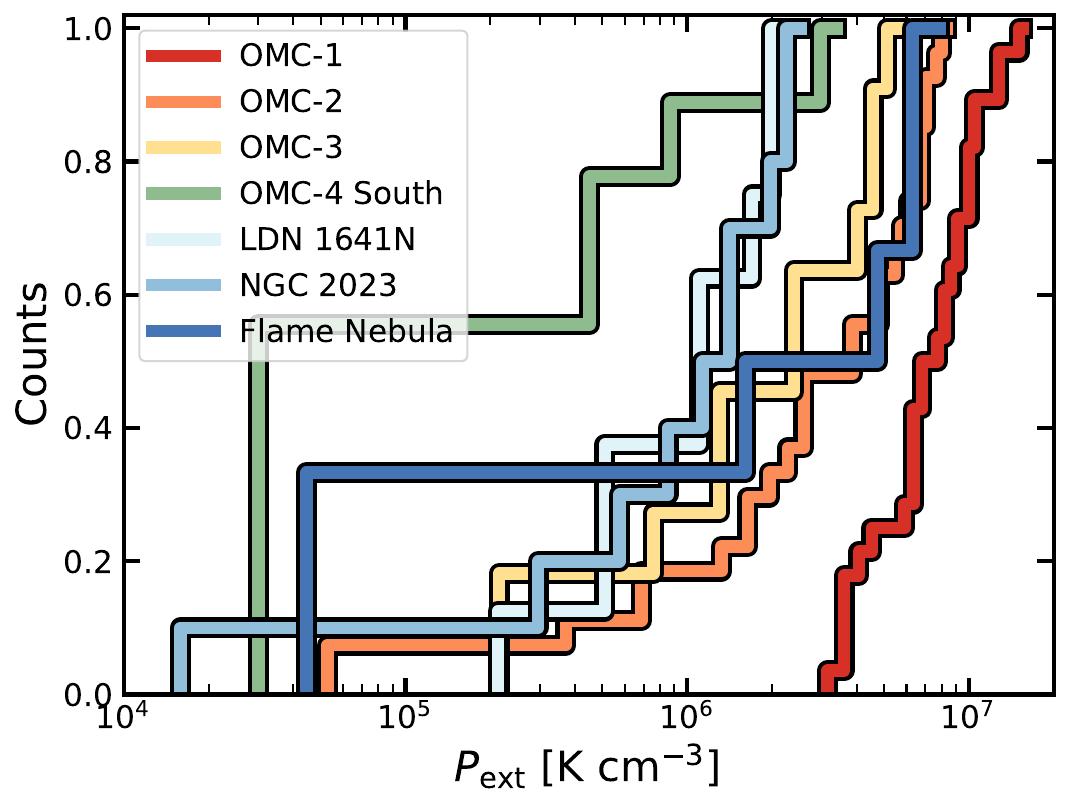}
\caption{Cumulative distributions of the external pressures ($P_\mathrm{ext}$) estimated via Eq.~(\ref{eq:Psurface}) and colour-coded per region.}
\label{fig:pext}
\end{figure}

Beyond the scatter seen in Fig.~\ref{fig:filcompar}, which could be ascribed to different conditions within regions, we speculate that the systematic variation in the fiber widths within our survey may depend not only on the peak column density ($N_0$), but also on the central gas volume density ($n_0$; see discussion in Sect.~\ref{subsec:fwhmn0dep}). From \citet{fischera12}, we give a broad estimate of this volume density for massive filaments ($f_\mathrm{cyl}=m/m_\mathrm{crit}\gtrsim0.6$, or $N_0\gtrsim10^{22}$~cm$^{-2}$). Assuming our fibers as isothermal, sub-critical, and pressure-confined filaments, the mean volume density can be determined in a first-order approximation as \citep[see Eqs.19 and 23 in][]{fischera12}
\begin{equation}\label{eq:fischera}
    n_0\sim \frac{N_0}{FWHM}.
\end{equation}
The combined effect of decreasing width and increasing peak column density seen in Fig.~\ref{fig:filcompar} could be linked to an increase in volume density across the survey. By computing this estimate per fiber, we obtain values within $n\sim10^5-10^6$~cm$^{-3}$. These values are above the density regime traced by N$_2$H$^+$ (1$-$0) \citep{bergin07}, and represent a reasonable lower limit for the volume densities previously estimated in some of our targets with higher-J transitions of N$_2$H$^+$ \citep[e.g. $\sim10^7$~cm$^{-3}$ in OMC-1;][]{teng20,hacar20sepia}. Despite Eq.~(\ref{eq:fischera}) represents only a first guess, it is also a promising hint towards the density, both column and volume, being the main factor driving the fiber widths.

Together with the above density estimates, the sub-virial line masses reported in our survey ($m/m_\mathrm{vir}<1$; see Table~\ref{tab:gen_prop}) suggest the presence of an external pressure $P_\mathrm{ext}$ in order to balance the high internal thermal pressure expected for the Orion fibers $P_0/k_\mathrm{B}=n_0 T_\mathrm{K}^0$.
The ratio between the inner $<P>$ and outer $P_\mathrm{ext}$ pressures in isothermal, pressure-truncated (and non-magnetised) filaments is expected to follow a linear relation \citep{Fiege2000}
\begin{equation}\label{eq:Pratio}
    \frac{P_\mathrm{ext}}{<P>}=1-\frac{m}{m_\mathrm{vir}}.
\end{equation}
Adopting $<P>\sim P_0/k_\mathrm{B}=n_0 T_\mathrm{K}^0$ and substituting for density from Eq.~(\ref{eq:fischera}), we can then estimate the value of $P_\mathrm{ext}$ in our fibers as
\begin{equation}\label{eq:Psurface}
    P_\mathrm{ext}/k_\mathrm{B}\simeq\left(1-\frac{m}{m_\mathrm{vir}}\right)\frac{N_0}{FWHM}\ T_\mathrm{K}^0.
\end{equation}
Applied to those sub-virial fibers in our sample (99 targets), Eq.~(\ref{eq:Psurface}) retrieves mean external pressures between $P_\mathrm{ext}/k_\mathrm{B}\sim 8\times10^6$~K~cm$^{-3}$ in OMC-1 and $\sim 4\times10^5$~K~cm$^{-3}$ in OMC-4 South (see Table~\ref{tab:gen_prop}).
These values are at least an order of magnitude higher compared to the standard ISM pressures of $P_\mathrm{ext}/k_\mathrm{B}=2\times 10^4$~K~cm$^{-3}$ \citep{cox05}. Interestingly, high-mass star-forming regions such as OMC-1 and the Flame Nebula present systematically higher  $P_\mathrm{ext}/k_\mathrm{B}$ values than those derived in low-mass clouds such as OMC-4 South or NGC~2023 (see Fig.~\ref{fig:pext}). 

The potential origin of these high external pressures is unclear and may have multiple sources. The warm diffuse environment found in certain regions in Orion such OMC-1 and the Flame Nebula, with background temperatures of $T_\mathrm{bg}\gtrsim30$~K, can contribute with addition thermal pressure $P_\mathrm{ext}^\mathrm{th}$. Yet, this thermal component would require high background densities of $n_\mathrm{bg}\gtrsim 10^4$~cm$^{-3}$ for $P_\mathrm{ext}^\mathrm{th}/k_\mathrm{B}=n_\mathrm{bg} T_{\rm{bg}}$ to match the previously estimated pressure values. Alternatively, infall and/or turbulent accretion (parameterised by an external velocity dispersion $\sigma_\mathrm{ext}$) can exert additional ram pressure $P_\mathrm{ext}^\mathrm{ram}=n_\mathrm{bg} \times \sigma_\mathrm{ext}^2$ \citep{heitsch13_acc,heigl2018} to support this configuration (i.e. $P_\mathrm{ext}=P_\mathrm{ext}^\mathrm{th}+P_\mathrm{ext}^\mathrm{ram}$).

Although informative about the physical conditions within and around fibers, the above first-order calculations assume an equilibrium solution for these objects. The non-isothermal nature of the Orion fibers suggests however either a more complex equilibrium condition \citep[e.g.][]{toci15}, or no equilibrium at all. Filaments and fibers dynamically evolving with their surrounding environment are to be considered. Additional independent measurements of the various energy contributions within and onto fiber are thus needed to investigate their possible equilibrium.

The analysis of our EMERGE Early ALMA Survey demonstrates a systematic variation in the fiber widths and column density in regions of different mass and complexity (Sects.~\ref{sec:fiberFWHM}, \ref{subsec:systvar}). Our results suggest a direct dependence of these two fiber properties in connection to the physical conditions of the cloud, in particular the external gas pressure and gas density (this section). Our findings suggest a smooth column density scaling between the different star formation regimes explored by our survey, in which fibers present a different $FWHM$ depending on their local densities and pressures.

\section{Conclusions}\label{sec:conclusions}

We have characterised here the radial properties of the fibers explored by the EMERGE Early ALMA Survey in Orion (see Paper III). The 152 objects identified in the seven star-forming regions of the survey describe the dense gas organisation at 4.5$''$ (or 2000~au) resolution sampling different star formation regimes, cloud morphology, and evolutionary stages. By exploring the gas column density and  kinetic temperature radial profiles of these fibers, we determined the main results listed below:

\begin{enumerate}
    \item Fibers show sharp column density profiles varying by more than two orders of magnitude within $<0.2$~pc in radius. These fiber profiles are well described by a single-peaked column density distribution, and reliably reproduced by a Gaussian fit. Using FilChap (Sect.~\ref{subsec:oldvsnew}), we obtained 991 individual fiber profiles allowing us to explore the peak column densities ($N_0$), widths ($FWHM$) and temperature gradients ($\nabla T_\mathrm{K}$) throughout a significant sample of targets.
    
    \item Fibers show column density profiles with a width within $\sim0.01-0.2$~pc, and a corresponding median value of 0.05~pc. More than 95\% of the radial cuts analysed showed $FWHM$ values below the typical filament width of $\sim 0.1$~pc reported in different {\it Herschel}-based studies (Sect.~\ref{sec:fiberFWHM}). 

    \item The analysis of the column density profiles in our targets returned a median peak column density of $N_0\sim 5\times 10^{22}$~cm$^{-2}$ ($\sim 50$~A$_\mathrm{V}$) and peak values up to $\gtrsim10^{23}$~cm$^{-2}$ ($> 100$~A$_\mathrm{V}$) in regions such as OMC-1 and the Flame Nebula (Sect.~\ref{sec:fiberN0}). 
    The typical peak column density of the Orion fibers is an order of magnitude higher than the peak column densities previously reported for parsec-scale filaments using {\it Herschel} observations.

    \item The vast majority of fibers in Orion clearly depart from the isothermal conditions usually assumed for filaments. By analysing the temperature radial profiles, we estimated typical radial temperature gradients above $>30$~K~pc$^{-1}$ in these fibers, with a corresponding median for the survey of $\sim80$~K~pc$^{-1}$. These temperature gradients are an order of magnitude larger than those determined in filaments within low-mass clouds such as Taurus or Musca. 

    \item We observed a systematic variation in the fiber $FWHM$ across the regions within the EMERGE Early ALMA Survey. On average, fibers in low-mass star-forming regions, such as NGC~2023 or OMC-4~South, are broader ($\sim0.06-0.07$~pc) than those found in high-mass star-forming regions, such as OMC-1 or the Flame Nebula ($\sim0.03-0.04$~pc) (Sect.~\ref{sec:theorobs}). Our ALMA survey provides the first robust evidence of a systematic change in the fiber widths depending on the environment, challenging the existence of a characteristic width for filaments. 

    \item We identified a direct anti-correlation between the observe fiber $FWHM$ and their peak column density, $N_0$ (Sect.~\ref{subsec:coldensitydep}). This $FWHM-N_0$ anti-correlation was predicted by different theoretical models for filaments with $N_0>10^{22}$~cm$^{-2}$, yet not detected by previous surveys in low-mass star-forming regions. Although the applicability of these models is challenged by the properties of the Orion fibers (mostly non-isothermal and prolate spheroids), the $FWHM-N_0$ dependence becomes evident for the column density regime these fibers probe. 

    \item We speculate that the reported $FWHM-N_0$ anti-correlation in the Orion fibers may result from the combined effect of the high internal gas densities and large external pressures estimated for the dense gas in these active star-forming regions (Sect.~\ref{subsec:speculation}).

    \item Our results suggest that fibers may adapt their radial profile and $FWHM$ to the environmental conditions; however, additional observations are required to constrain the external gas conditions and the dynamical state of fibers.

\end{enumerate}

\begin{acknowledgements}
    The authors thank the anonymous referee for the extensive review provided which helped improving the manuscript.
    This project has received funding from the European Research Council (ERC) under the European Union’s Horizon 2020 research and innovation programme (Grant agreement No. 851435).
    M.T. acknowledges partial support from project PID2019-108765GB-I00 funded by MCIN/AEI/10.13039/501100011033.
    This paper makes use of the following ALMA data: ADS/JAO.ALMA\#2019.1.00641.S., ADS/JAO.ALMA\#2015.1.00669.S. ALMA is a partnership of ESO (representing its member states), NSF (USA) and NINS (Japan), together with NRC (Canada), MOST and ASIAA (Taiwan), and KASI (Republic of Korea), in cooperation with the Republic of Chile. The Joint ALMA Observatory is operated by ESO, AUI/NRAO and NAOJ.
    This work is based on IRAM-30m telescope observations carried out under project numbers 032-13, 120-20, 060-22, and 133-22. IRAM is supported by INSU/CNRS (France), MPG (Germany), and IGN (Spain). 
    This research has made use of the SIMBAD database, operated at CDS, Strasbourg, France.
    This research has made use of NASA’s Astrophysics Data System.
\end{acknowledgements}

\nocite{*}
\bibliographystyle{aa}
\bibliography{bibl.bib}

\begin{thebibliography}{71}
\expandafter\ifx\csname natexlab\endcsname\relax\def\natexlab#1{#1}\fi

\bibitem[{{Andr{\'e}} {et~al.}(2014){Andr{\'e}}, {Di Francesco}, {Ward-Thompson}, {Inutsuka}, {Pudritz}, \& {Pineda}}]{andre14}
{Andr{\'e}}, P., {Di Francesco}, J., {Ward-Thompson}, D., {et~al.} 2014, in Protostars and Planets VI, ed. H.~{Beuther}, R.~S. {Klessen}, C.~P. {Dullemond}, \& T.~{Henning}, 27--51

\bibitem[{{Andr{\'e}} {et~al.}(2010){Andr{\'e}}, {Men'shchikov}, {Bontemps}, {K{\"o}nyves}, {Motte}, {Schneider}, {Didelon}, {Minier}, {Saraceno}, {Ward-Thompson}, {di Francesco}, {White}, {Molinari}, {Testi}, {Abergel}, {Griffin}, {Henning}, {Royer}, {Mer{\'\i}n}, {Vavrek}, {Attard}, {Arzoumanian}, {Wilson}, {Ade}, {Aussel}, {Baluteau}, {Benedettini}, {Bernard}, {Blommaert}, {Cambr{\'e}sy}, {Cox}, {di Giorgio}, {Hargrave}, {Hennemann}, {Huang}, {Kirk}, {Krause}, {Launhardt}, {Leeks}, {Le Pennec}, {Li}, {Martin}, {Maury}, {Olofsson}, {Omont}, {Peretto}, {Pezzuto}, {Prusti}, {Roussel}, {Russeil}, {Sauvage}, {Sibthorpe}, {Sicilia-Aguilar}, {Spinoglio}, {Waelkens}, {Woodcraft}, \& {Zavagno}}]{andre10}
{Andr{\'e}}, P., {Men'shchikov}, A., {Bontemps}, S., {et~al.} 2010, \aap, 518, L102

\bibitem[{{Andr{\'e}} {et~al.}(2022){Andr{\'e}}, {Palmeirim}, \& {Arzoumanian}}]{andre22}
{Andr{\'e}}, P.~J., {Palmeirim}, P., \& {Arzoumanian}, D. 2022, \aap, 667, L1

\bibitem[{{Arzoumanian} {et~al.}(2011){Arzoumanian}, {Andr{\'e}}, {Didelon}, {K{\"o}nyves}, {Schneider}, {Men'shchikov}, {Sousbie}, {Zavagno}, {Bontemps}, {di Francesco}, {Griffin}, {Hennemann}, {Hill}, {Kirk}, {Martin}, {Minier}, {Molinari}, {Motte}, {Peretto}, {Pezzuto}, {Spinoglio}, {Ward-Thompson}, {White}, \& {Wilson}}]{arzoumanian11}
{Arzoumanian}, D., {Andr{\'e}}, P., {Didelon}, P., {et~al.} 2011, \aap, 529, L6

\bibitem[{{Arzoumanian} {et~al.}(2019){Arzoumanian}, {Andr{\'e}}, {K{\"o}nyves}, {Palmeirim}, {Roy}, {Schneider}, {Benedettini}, {Didelon}, {Di Francesco}, {Kirk}, \& {Ladjelate}}]{arzoumanian19}
{Arzoumanian}, D., {Andr{\'e}}, P., {K{\"o}nyves}, V., {et~al.} 2019, \aap, 621, A42

\bibitem[{{Asplund} {et~al.}(2021){Asplund}, {Amarsi}, \& {Grevesse}}]{aspl21}
{Asplund}, M., {Amarsi}, A.~M., \& {Grevesse}, N. 2021, \aap, 653, A141

\bibitem[{{Barnard}(1907)}]{barnard07}
{Barnard}, E.~E. 1907, \apj, 25, 218

\bibitem[{{Bergin} \& {Tafalla}(2007)}]{bergin07}
{Bergin}, E.~A. \& {Tafalla}, M. 2007, \araa, 45, 339

\bibitem[{{Bonanomi} {et~al.}(2024){Bonanomi}, {Hacar}, {Socci}, {Petry}, \& {Suri}}]{bonanomi2024}
{Bonanomi}, F., {Hacar}, A., {Socci}, A., {Petry}, D., \& {Suri}, S. 2024, \aap, 688, A30

\bibitem[{{Bonne} {et~al.}(2020){Bonne}, {Bontemps}, {Schneider}, {Clarke}, {Arzoumanian}, {Fukui}, {Tachihara}, {Csengeri}, {Guesten}, {Ohama}, {Okamoto}, {Simon}, {Yahia}, \& {Yamamoto}}]{bonne20}
{Bonne}, L., {Bontemps}, S., {Schneider}, N., {et~al.} 2020, \aap, 644, A27

\bibitem[{{Cox}(2005)}]{cox05}
{Cox}, D.~P. 2005, \araa, 43, 337

\bibitem[{{Dhabal} {et~al.}(2019){Dhabal}, {Mundy}, {Chen}, {Teuben}, \& {Storm}}]{dhab19}
{Dhabal}, A., {Mundy}, L.~G., {Chen}, C.-y., {Teuben}, P., \& {Storm}, S. 2019, \apj, 876, 108

\bibitem[{{Federrath}(2016)}]{federrath16}
{Federrath}, C. 2016, \mnras, 457, 375

\bibitem[{{Fern{\'a}ndez-L{\'o}pez} {et~al.}(2014){Fern{\'a}ndez-L{\'o}pez}, {Arce}, {Looney}, {Mundy}, {Storm}, {Teuben}, {Lee}, {Segura-Cox}, {Isella}, {Tobin}, {Rosolowsky}, {Plunkett}, {Kwon}, {Kauffmann}, {Ostriker}, {Tassis}, {Shirley}, \& {Pound}}]{fernandez14}
{Fern{\'a}ndez-L{\'o}pez}, M., {Arce}, H.~G., {Looney}, L., {et~al.} 2014, \apjl, 790, L19

\bibitem[{{Fiege} \& {Pudritz}(2000)}]{Fiege2000}
{Fiege}, J.~D. \& {Pudritz}, R.~E. 2000, \mnras, 311, 85

\bibitem[{{Fischera} \& {Martin}(2012)}]{fischera12}
{Fischera}, J. \& {Martin}, P.~G. 2012, \aap, 542, A77

\bibitem[{{Furlan} {et~al.}(2016){Furlan}, {Fischer}, {Ali}, {Stutz}, {Stanke}, {Tobin}, {Megeath}, {Osorio}, {Hartmann}, {Calvet}, {Poteet}, {Booker}, {Manoj}, {Watson}, \& {Allen}}]{furlan16}
{Furlan}, E., {Fischer}, W.~J., {Ali}, B., {et~al.} 2016, \apjs, 224, 5

\bibitem[{{Gehman} {et~al.}(1996){Gehman}, {Adams}, {Fatuzzo}, \& {Watkins}}]{Gehman1996}
{Gehman}, C.~S., {Adams}, F.~C., {Fatuzzo}, M., \& {Watkins}, R. 1996, \apj, 457, 718

\bibitem[{{Goldsmith}(2001)}]{goldsmith2001}
{Goldsmith}, P.~F. 2001, \apj, 557, 736

\bibitem[{{Goodman} {et~al.}(2014){Goodman}, {Alves}, {Beaumont}, {Benjamin}, {Borkin}, {Burkert}, {Dame}, {Jackson}, {Kauffmann}, {Robitaille}, \& {Smith}}]{goodman14}
{Goodman}, A.~A., {Alves}, J., {Beaumont}, C.~N., {et~al.} 2014, \apj, 797, 53

\bibitem[{{Hacar} {et~al.}(2017{\natexlab{a}}){Hacar}, {Alves}, {Tafalla}, \& {Goicoechea}}]{hacar17b}
{Hacar}, A., {Alves}, J., {Tafalla}, M., \& {Goicoechea}, J.~R. 2017{\natexlab{a}}, \aap, 602, L2

\bibitem[{{Hacar} {et~al.}(2023){Hacar}, {Clark}, {Heitsch}, {Kainulainen}, {Panopoulou}, {Seifried}, \& {Smith}}]{hacar22}
{Hacar}, A., {Clark}, S.~E., {Heitsch}, F., {et~al.} 2023, in Astronomical Society of the Pacific Conference Series, Vol. 534, Protostars and Planets VII, ed. S.~{Inutsuka}, Y.~{Aikawa}, T.~{Muto}, K.~{Tomida}, \& M.~{Tamura}, 153

\bibitem[{{Hacar} {et~al.}(2020){Hacar}, {Hogerheijde}, {Harsono}, {Portegies Zwart}, {De Breuck}, {Torstensson}, {Boland}, {Baryshev}, {Hesper}, {Barkhof}, {Adema}, {Bekema}, {Koops}, {Khudchenko}, \& {Stark}}]{hacar20sepia}
{Hacar}, A., {Hogerheijde}, M.~R., {Harsono}, D., {et~al.} 2020, \aap, 644, A133

\bibitem[{{Hacar} {et~al.}(2024){Hacar}, {Socci}, {Bonanomi}, {Petry}, {Tafalla}, {Harsono}, {Forbrich}, {Alves}, {Grossschedl}, {Goicoechea}, {Pety}, {Burkert}, \& {Li}}]{hacar24}
{Hacar}, A., {Socci}, A., {Bonanomi}, F., {et~al.} 2024, \aap, 687, A140

\bibitem[{{Hacar} \& {Tafalla}(2011)}]{hacar11}
{Hacar}, A. \& {Tafalla}, M. 2011, \aap, 533, A34

\bibitem[{{Hacar} {et~al.}(2017{\natexlab{b}}){Hacar}, {Tafalla}, \& {Alves}}]{hacar17}
{Hacar}, A., {Tafalla}, M., \& {Alves}, J. 2017{\natexlab{b}}, \aap, 606, A123

\bibitem[{{Hacar} {et~al.}(2018){Hacar}, {Tafalla}, {Forbrich}, {Alves}, {Meingast}, {Grossschedl}, \& {Teixeira}}]{hacar18}
{Hacar}, A., {Tafalla}, M., {Forbrich}, J., {et~al.} 2018, \aap, 610, A77

\bibitem[{{Hacar} {et~al.}(2013){Hacar}, {Tafalla}, {Kauffmann}, \& {Kov{\'a}cs}}]{hacar13}
{Hacar}, A., {Tafalla}, M., {Kauffmann}, J., \& {Kov{\'a}cs}, A. 2013, \aap, 554, A55

\bibitem[{{Hanawa} {et~al.}(1993){Hanawa}, {Nakamura}, {Matsumoto}, {Nakano}, {Tatematsu}, {Umemoto}, {Kameya}, {Hirano}, {Hasegawa}, {Kaifu}, \& {Yamamoto}}]{hanawa1993}
{Hanawa}, T., {Nakamura}, F., {Matsumoto}, T., {et~al.} 1993, \apjl, 404, L83

\bibitem[{{Hartmann}(2002)}]{hartmann02}
{Hartmann}, L. 2002, \apj, 578, 914

\bibitem[{{Heigl} {et~al.}(2018){Heigl}, {Burkert}, \& {Gritschneder}}]{heigl2018}
{Heigl}, S., {Burkert}, A., \& {Gritschneder}, M. 2018, \mnras, 474, 4881

\bibitem[{{Heitsch}(2013{\natexlab{a}})}]{heitsch13_acc}
{Heitsch}, F. 2013{\natexlab{a}}, \apj, 769, 115

\bibitem[{{Heitsch}(2013{\natexlab{b}})}]{heitsch13}
{Heitsch}, F. 2013{\natexlab{b}}, \apj, 776, 62

\bibitem[{{Hennemann} {et~al.}(2012){Hennemann}, {Motte}, {Schneider}, {Didelon}, {Hill}, {Arzoumanian}, {Bontemps}, {Csengeri}, {Andr{\'e}}, {Konyves}, {Louvet}, {Marston}, {Men'shchikov}, {Minier}, {Nguyen Luong}, {Palmeirim}, {Peretto}, {Sauvage}, {Zavagno}, {Anderson}, {Bernard}, {Di Francesco}, {Elia}, {Li}, {Martin}, {Molinari}, {Pezzuto}, {Russeil}, {Rygl}, {Schisano}, {Spinoglio}, {Sousbie}, {Ward-Thompson}, \& {White}}]{hennemann12}
{Hennemann}, M., {Motte}, F., {Schneider}, N., {et~al.} 2012, \aap, 543, L3

\bibitem[{{Howard} {et~al.}(2019){Howard}, {Whitworth}, {Marsh}, {Clarke}, {Griffin}, {Smith}, \& {Lomax}}]{howard19}
{Howard}, A.~D.~P., {Whitworth}, A.~P., {Marsh}, K.~A., {et~al.} 2019, \mnras, 489, 962

\bibitem[{{Jackson} {et~al.}(2010){Jackson}, {Finn}, {Chambers}, {Rathborne}, \& {Simon}}]{jackson10}
{Jackson}, J.~M., {Finn}, S.~C., {Chambers}, E.~T., {Rathborne}, J.~M., \& {Simon}, R. 2010, \apjl, 719, L185

\bibitem[{{Johnstone} \& {Bally}(1999)}]{john99}
{Johnstone}, D. \& {Bally}, J. 1999, \apjl, 510, L49

\bibitem[{{Juvela} \& {Mannfors}(2023)}]{juve23}
{Juvela}, M. \& {Mannfors}, E. 2023, \aap, 671, A111

\bibitem[{{Kauffmann} {et~al.}(2008){Kauffmann}, {Bertoldi}, {Bourke}, {Evans}, \& {Lee}}]{kauff08}
{Kauffmann}, J., {Bertoldi}, F., {Bourke}, T.~L., {Evans}, N.~J., I., \& {Lee}, C.~W. 2008, \aap, 487, 993

\bibitem[{{Kawachi} \& {Hanawa}(1998)}]{kawachi1998}
{Kawachi}, T. \& {Hanawa}, T. 1998, \pasj, 50, 577

\bibitem[{{Lee} {et~al.}(2014){Lee}, {Fern{\'a}ndez-L{\'o}pez}, {Storm}, {Looney}, {Mundy}, {Segura-Cox}, {Teuben}, {Rosolowsky}, {Arce}, {Ostriker}, {Shirley}, {Kwon}, {Kauffmann}, {Tobin}, {Plunkett}, {Pound}, {Salter}, {Volgenau}, {Chen}, {Tassis}, {Isella}, {Crutcher}, {Gammie}, \& {Testi}}]{lee14}
{Lee}, K.~I., {Fern{\'a}ndez-L{\'o}pez}, M., {Storm}, S., {et~al.} 2014, \apj, 797, 76

\bibitem[{{Li} {et~al.}(2022){Li}, {Sanhueza}, {Lee}, {Zhang}, {Beuther}, {Palau}, {Liu}, {Smith}, {Liu}, {Jim{\'e}nez-Serra}, {Kim}, {Feng}, {Liu}, {Wang}, {Li}, {Qiu}, {Lu}, {Girart}, {Wang}, {Li}, {Li}, {Cao}, {Kim}, \& {Strom}}]{li22}
{Li}, S., {Sanhueza}, P., {Lee}, C.~W., {et~al.} 2022, \apj, 926, 165

\bibitem[{{Megeath} {et~al.}(2012){Megeath}, {Gutermuth}, {Muzerolle}, {Kryukova}, {Flaherty}, {Hora}, {Allen}, {Hartmann}, {Myers}, {Pipher}, {Stauffer}, {Young}, \& {Fazio}}]{megeath12}
{Megeath}, S.~T., {Gutermuth}, R., {Muzerolle}, J., {et~al.} 2012, \aj, 144, 192

\bibitem[{{Menten} {et~al.}(2007){Menten}, {Reid}, {Forbrich}, \& {Brunthaler}}]{menten07}
{Menten}, K.~M., {Reid}, M.~J., {Forbrich}, J., \& {Brunthaler}, A. 2007, \aap, 474, 515

\bibitem[{{Molinari} {et~al.}(2010){Molinari}, {Swinyard}, {Bally}, {Barlow}, {Bernard}, {Martin}, {Moore}, {Noriega-Crespo}, {Plume}, {Testi}, {Zavagno}, {Abergel}, {Ali}, {Anderson}, {Andr{\'e}}, {Baluteau}, {Battersby}, {Beltr{\'a}n}, {Benedettini}, {Billot}, {Blommaert}, {Bontemps}, {Boulanger}, {Brand}, {Brunt}, {Burton}, {Calzoletti}, {Carey}, {Caselli}, {Cesaroni}, {Cernicharo}, {Chakrabarti}, {Chrysostomou}, {Cohen}, {Compiegne}, {de Bernardis}, {de Gasperis}, {di Giorgio}, {Elia}, {Faustini}, {Flagey}, {Fukui}, {Fuller}, {Ganga}, {Garcia-Lario}, {Glenn}, {Goldsmith}, {Griffin}, {Hoare}, {Huang}, {Ikhenaode}, {Joblin}, {Joncas}, {Juvela}, {Kirk}, {Lagache}, {Li}, {Lim}, {Lord}, {Marengo}, {Marshall}, {Masi}, {Massi}, {Matsuura}, {Minier}, {Miville-Desch{\^e}nes}, {Montier}, {Morgan}, {Motte}, {Mottram}, {M{\"u}ller}, {Natoli}, {Neves}, {Olmi}, {Paladini}, {Paradis}, {Parsons}, {Peretto}, {Pestalozzi}, {Pezzuto}, {Piacentini}, {Piazzo}, {Polychroni}, {Pomar{\`e}s}, {Popescu}, {Reach}, {Ristorcelli},
  {Robitaille}, {Robitaille}, {Rod{\'o}n}, {Roy}, {Royer}, {Russeil}, {Saraceno}, {Sauvage}, {Schilke}, {Schisano}, {Schneider}, {Schuller}, {Schulz}, {Sibthorpe}, {Smith}, {Smith}, {Spinoglio}, {Stamatellos}, {Strafella}, {Stringfellow}, {Sturm}, {Taylor}, {Thompson}, {Traficante}, {Tuffs}, {Umana}, {Valenziano}, {Vavrek}, {Veneziani}, {Viti}, {Waelkens}, {Ward-Thompson}, {White}, {Wilcock}, {Wyrowski}, {Yorke}, \& {Zhang}}]{molinari10}
{Molinari}, S., {Swinyard}, B., {Bally}, J., {et~al.} 2010, \aap, 518, L100

\bibitem[{{Monsch} {et~al.}(2018){Monsch}, {Pineda}, {Liu}, {Zucker}, {How-Huan Chen}, {Pattle}, {Offner}, {Di Francesco}, {Ginsburg}, {Ercolano}, {Arce}, {Friesen}, {Kirk}, {Caselli}, \& {Goodman}}]{monsch18}
{Monsch}, K., {Pineda}, J.~E., {Liu}, H.~B., {et~al.} 2018, \apj, 861, 77

\bibitem[{{Nakamura} {et~al.}(1993){Nakamura}, {Hanawa}, \& {Nakano}}]{nakamura1993}
{Nakamura}, F., {Hanawa}, T., \& {Nakano}, T. 1993, \pasj, 45, 551

\bibitem[{{Ostriker}(1964)}]{ostriker64}
{Ostriker}, J. 1964, \apj, 140, 1056

\bibitem[{{Padoan} {et~al.}(2001){Padoan}, {Juvela}, {Goodman}, \& {Nordlund}}]{padoan01}
{Padoan}, P., {Juvela}, M., {Goodman}, A.~A., \& {Nordlund}, {\r{A}}. 2001, \apj, 553, 227

\bibitem[{{Palmeirim} {et~al.}(2013){Palmeirim}, {Andr{\'e}}, {Kirk}, {Ward-Thompson}, {Arzoumanian}, {K{\"o}nyves}, {Didelon}, {Schneider}, {Benedettini}, {Bontemps}, {Di Francesco}, {Elia}, {Griffin}, {Hennemann}, {Hill}, {Martin}, {Men'shchikov}, {Molinari}, {Motte}, {Nguyen Luong}, {Nutter}, {Peretto}, {Pezzuto}, {Roy}, {Rygl}, {Spinoglio}, \& {White}}]{palme13}
{Palmeirim}, P., {Andr{\'e}}, P., {Kirk}, J., {et~al.} 2013, \aap, 550, A38

\bibitem[{{Panopoulou} {et~al.}(2022){Panopoulou}, {Clark}, {Hacar}, {Heitsch}, {Kainulainen}, {Ntormousi}, {Seifried}, \& {Smith}}]{pano22}
{Panopoulou}, G.~V., {Clark}, S.~E., {Hacar}, A., {et~al.} 2022, \aap, 657, L13

\bibitem[{{Panopoulou} {et~al.}(2017){Panopoulou}, {Psaradaki}, {Skalidis}, {Tassis}, \& {Andrews}}]{panopoulou2017}
{Panopoulou}, G.~V., {Psaradaki}, I., {Skalidis}, R., {Tassis}, K., \& {Andrews}, J.~J. 2017, \mnras, 466, 2529

\bibitem[{{Panopoulou} {et~al.}(2014){Panopoulou}, {Tassis}, {Goldsmith}, \& {Heyer}}]{pano14}
{Panopoulou}, G.~V., {Tassis}, K., {Goldsmith}, P.~F., \& {Heyer}, M.~H. 2014, \mnras, 444, 2507

\bibitem[{{Pineda} {et~al.}(2023){Pineda}, {Arzoumanian}, {Andre}, {Friesen}, {Zavagno}, {Clarke}, {Inoue}, {Chen}, {Lee}, {Soler}, \& {Kuffmeier}}]{pineda23}
{Pineda}, J.~E., {Arzoumanian}, D., {Andre}, P., {et~al.} 2023, in Astronomical Society of the Pacific Conference Series, Vol. 534, Protostars and Planets VII, ed. S.~{Inutsuka}, Y.~{Aikawa}, T.~{Muto}, K.~{Tomida}, \& M.~{Tamura}, 233

\bibitem[{{Pineda} {et~al.}(2011){Pineda}, {Goodman}, {Arce}, {Caselli}, {Longmore}, \& {Corder}}]{pineda2011}
{Pineda}, J.~E., {Goodman}, A.~A., {Arce}, H.~G., {et~al.} 2011, \apjl, 739, L2

\bibitem[{{Priestley} {et~al.}(2023){Priestley}, {Arzoumanian}, \& {Whitworth}}]{Priestley2023}
{Priestley}, F.~D., {Arzoumanian}, D., \& {Whitworth}, A.~P. 2023, \mnras, 522, 3890

\bibitem[{{Priestley} \& {Whitworth}(2022)}]{priest22}
{Priestley}, F.~D. \& {Whitworth}, A.~P. 2022, \mnras, 512, 1407

\bibitem[{{Recchi} {et~al.}(2013){Recchi}, {Hacar}, \& {Palestini}}]{recchi13}
{Recchi}, S., {Hacar}, A., \& {Palestini}, A. 2013, \aap, 558, A27

\bibitem[{{Recchi} {et~al.}(2014){Recchi}, {Hacar}, \& {Palestini}}]{recchi2014}
{Recchi}, S., {Hacar}, A., \& {Palestini}, A. 2014, \mnras, 444, 1775

\bibitem[{{Schmiedeke} {et~al.}(2021){Schmiedeke}, {Pineda}, {Caselli}, {Arce}, {Fuller}, {Goodman}, {Maureira}, {Offner}, {Segura-Cox}, \& {Seifried}}]{schmi21}
{Schmiedeke}, A., {Pineda}, J.~E., {Caselli}, P., {et~al.} 2021, \apj, 909, 60

\bibitem[{{Schuller} {et~al.}(2021){Schuller}, {Andr{\'e}}, {Shimajiri}, {Zavagno}, {Peretto}, {Arzoumanian}, {Csengeri}, {K{\"o}nyves}, {Palmeirim}, {Pezzuto}, {Rigby}, {Roussel}, {Ajeddig}, {Dumaye}, {Gallais}, {Le Pennec}, {Martignac}, {Mattern}, {Rev{\'e}ret}, {Rodriguez}, \& {Talvard}}]{schuller21}
{Schuller}, F., {Andr{\'e}}, P., {Shimajiri}, Y., {et~al.} 2021, \aap, 651, A36

\bibitem[{{Socci} {et~al.}(2024){Socci}, {Hacar}, {Bonanomi}, {Tafalla}, \& {Suri}}]{socci24}
{Socci}, A., {Hacar}, A., {Bonanomi}, F., {Tafalla}, M., \& {Suri}, S. 2024, submitted

\bibitem[{{Stepnik} {et~al.}(2003){Stepnik}, {Abergel}, {Bernard}, {Boulanger}, {Cambr{\'e}sy}, {Giard}, {Jones}, {Lagache}, {Lamarre}, {Meny}, {Pajot}, {Le Peintre}, {Ristorcelli}, {Serra}, \& {Torre}}]{stepnik03}
{Stepnik}, B., {Abergel}, A., {Bernard}, J.~P., {et~al.} 2003, \aap, 398, 551

\bibitem[{{Stod{\'o}lkiewicz}(1963)}]{stodo63}
{Stod{\'o}lkiewicz}, J.~S. 1963, \actaa, 13, 30

\bibitem[{{Stutz} {et~al.}(2013){Stutz}, {Tobin}, {Stanke}, {Megeath}, {Fischer}, {Robitaille}, {Henning}, {Ali}, {di Francesco}, {Furlan}, {Hartmann}, {Osorio}, {Wilson}, {Allen}, {Krause}, \& {Manoj}}]{stutz13}
{Stutz}, A.~M., {Tobin}, J.~J., {Stanke}, T., {et~al.} 2013, \apj, 767, 36

\bibitem[{{Suri} {et~al.}(2019){Suri}, {S{\'a}nchez-Monge}, {Schilke}, {Clarke}, {Smith}, {Ossenkopf-Okada}, {Klessen}, {Padoan}, {Goldsmith}, {Arce}, {Bally}, {Carpenter}, {Ginsburg}, {Johnstone}, {Kauffmann}, {Kong}, {Lis}, {Mairs}, {Pillai}, {Pineda}, \& {Duarte-Cabral}}]{suri19}
{Suri}, S., {S{\'a}nchez-Monge}, {\'A}., {Schilke}, P., {et~al.} 2019, \aap, 623, A142

\bibitem[{{Tafalla} \& {Hacar}(2015)}]{tafalla15}
{Tafalla}, M. \& {Hacar}, A. 2015, \aap, 574, A104

\bibitem[{{Tafalla} {et~al.}(2023){Tafalla}, {Usero}, \& {Hacar}}]{tafalla23}
{Tafalla}, M., {Usero}, A., \& {Hacar}, A. 2023, \aap, 679, A112

\bibitem[{{Teng} \& {Hirano}(2020)}]{teng20}
{Teng}, Y.-H. \& {Hirano}, N. 2020, \apj, 893, 63

\bibitem[{{Toci} \& {Galli}(2015)}]{toci15}
{Toci}, C. \& {Galli}, D. 2015, \mnras, 446, 2110

\bibitem[{{Wenger} {et~al.}(2000){Wenger}, {Ochsenbein}, {Egret}, {Dubois}, {Bonnarel}, {Borde}, {Genova}, {Jasniewicz}, {Lalo{\"e}}, {Lesteven}, \& {Monier}}]{wenger00}
{Wenger}, M., {Ochsenbein}, F., {Egret}, D., {et~al.} 2000, \aaps, 143, 9

\end{thebibliography}



\begin{appendix}

\section{Integration of FilChap}\label{sec:filchap}

FilChap is an automatic fitting routine of radial profiles either in intensity or column density. In this section, we first discuss in depth the main changes made to the its public version to integrate it in our analysis (Sect.~\ref{subsec:intfilchap}). We then describe the tests performed to assess the best fitting function for our radial profiles using the parameters provided by FilChap (Sect.~\ref{subsec:symm}).

\subsection{Changes to the public version}\label{subsec:intfilchap}

In the following, we describe the changes made to the publicly available version of FilChap to adapt its outputs to our filament-finding algorithm:
\begin{itemize}
    \item \texttt{n2}: originally set to the same value of \texttt{avg$\_$len} \citep{suri19}, \texttt{n2} is the length of each fraction of the axis used to compute the average profile and perform the fitting. The process is then iterated over the whole axis with the output being the best fit parameters per fraction. Our algorithm already determines the axis knots as density weighted averages (A$_\mathrm{v}^2$) in cuts of size $2\times\theta_\mathrm{beam}$ \citep[see Paper III and][]{hacar13}. We therefore set \texttt{avg$\_$len} = 1 and \texttt{n2} = 2 to have the radial sampling performed only around these points;
    \item \texttt{npix}: size, in pixels, of the radial cut on each side of the axis point. Set to 45 and 43 pixels as equivalent of 0.2~pc at the distance of Orion A and B \citep[400~pc and 423~pc, respectively; see][for the full discussion]{hacar24} given a pixel size of 2.25\arcsec ($\theta_\mathrm{beam}/2$). The choice of 0.2~pc is a compromise between including too many empty pixels in the radial cut and avoiding biasing the fit towards narrower values by not including enough fields in the average;
    \item \texttt{dist$\_$normal}: width of the perpendicular sampling done on each side of the axis knots. The width of this stripe is set to $3\times\theta_\mathrm{beam}$, over which FilChap computes the average radial profile. Given a distance of $2\times\theta_\mathrm{beam}$ between axis knots, FilChap returns a good average profile at the expense of a minimum superposition;
    \item \texttt{intensity}: originally the intensity map, now this variable is instead the column density map sampled along the fiber axes (see Paper III and Fig.~\ref{fig:OMC3showcase}, central panel). For each of the resulting radial cuts, we excluded the baseline subtraction routine since we disregarded the floor term in the determination of N(H$_2$) (see Paper III);
    \item \texttt{intensity$\_$Tkin}: similarly to the \texttt{intensity} parameter, we provide an additional file which contains the temperature map (see Paper III and Fig.~\ref{fig:OMC3showcase}, right panel). The temperature map undergoes the same sampling of the column density map along the radial cut, but without any averaging over \texttt{dist$\_$normal}. Thus, we can qualitatively compare, although at different resolutions, the two profiles (see Figs.~\ref{fig:omc3structureprofiles},\ref{fig:radialprofilessur});
    \item \texttt{order}: originally set to 6 to adapt to the CARMA-NRO data \citep{suri19}, \texttt{order} is the number of pixels on each side of a extreme point to assess its first and second derivatives. Based on their values, FilChap determines whether it is a minimum, maximum or inflection point of the profile. We set the parameter to 3, and thus a total interval of 6 pixels around the point (i.e. $3\times\theta_\mathrm{beam}$);
    \item \texttt{limit}: originally defined in intensity, this parameter sets the limit to assess, along with the derivatives, whether a extreme point is a peak or a shoulder and equates $5\times\sigma$, where $\sigma$ is the user-defined noise. We convert the average noise determined across our maps ($\sim0.75$~K~km~s$^{-1}$) into column density using Eq.~(1) in Paper III obtaining $\sigma\sim3\times10^{21}$~cm$^{-2}$ for a typical temperature of 20~K seen across our fields.
\end{itemize}

\subsection{Peaks, shoulders, and asymmetries in the profiles}\label{subsec:symm}

\begin{figure*}
    \centering
    \includegraphics[width=\linewidth]{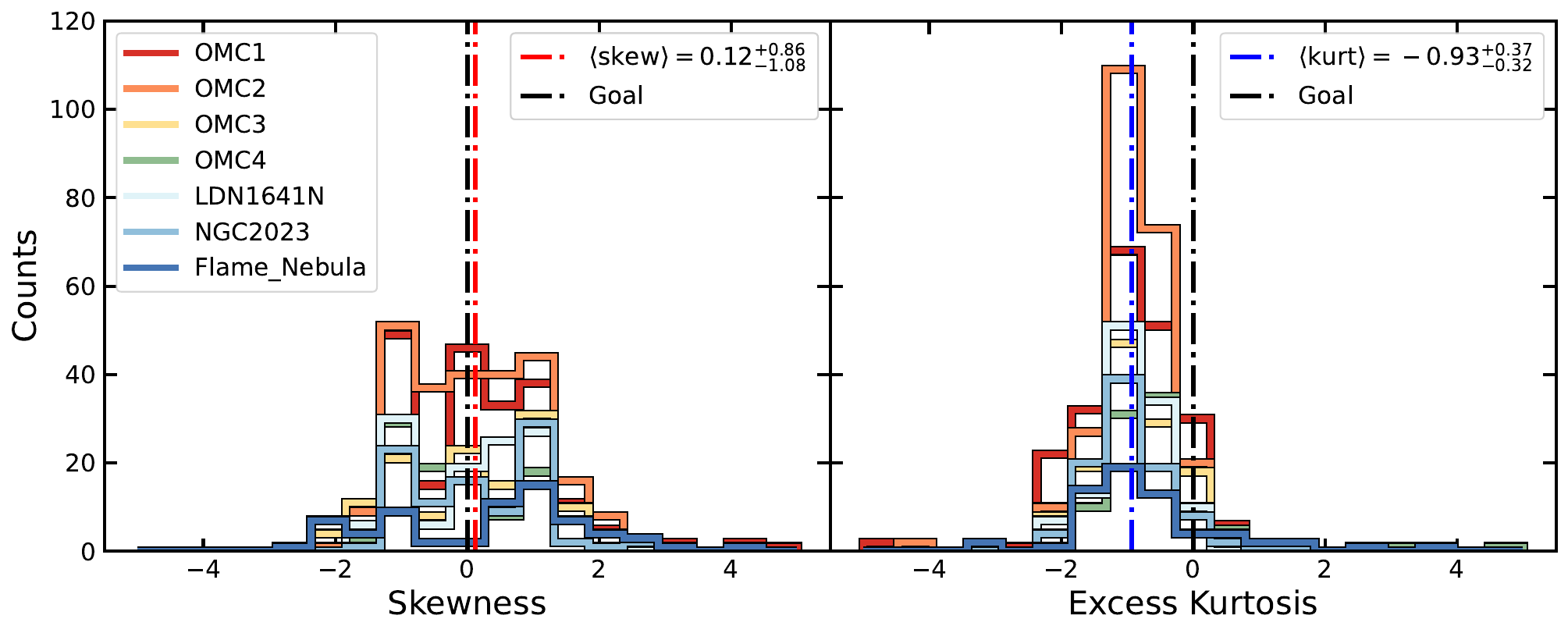}
    \caption{Skewness (\textbf{left panel}) and excess kurtosis (\textbf{right panel}) defined in Eq.~(\ref{eq:kurtskew}) and determined for the individual cuts in our survey. Both parameters are colour-coded per region, and vertical red and blue represent their median values. The black vertical line in both panels represents the expected value for a Gaussian profile.}
    \label{fig:kurtskew}
\end{figure*}

In this Section, we describe the features observed in the radial profiles using the parameters provided by FilChap. These features allowed us to quantitatively select the fitting function best suited for their analysis.

The description of the asymmetries, multiplicity of peaks, and high-contrast features of the radial profiles (see Figs.~\ref{fig:omc3structureprofiles}-~\ref{fig:radialprofilessur}) starts by exploring their $skewness$ and $(excess)~kurtosis$. The two parameters, describing the overall departure from a Gaussian profile, are defined within FilChap as follows:
\begin{equation}\label{eq:kurtskew}
    skew = \frac{\mu_3}{\sigma^3}, \quad kurt = \frac{\mu_4}{\sigma^4} - 3,
\end{equation}
where $\mu_i$ are the moments of the radial distribution and $\sigma$ is its standard deviation.

Figure~\ref{fig:kurtskew} shows the distributions of skewness (left panel) and excess kurtosis (right panel) for all the cuts in our survey. The skewness shows a median value of $0.12^{+0.86}_{-1.08}$, suggesting symmmetric profiles on average. However, when looking at the scatter in the distribution, the profiles are clearly more complex than a simple Gaussian, as is shown in Figs.~\ref{fig:omc3structureprofiles}-\ref{fig:radialprofilessur}. The excess kurtosis is instead peaked around $\sim-1$ with a low degree of scatter (the corresponding median value is $-0.93^{+0.37}_{-0.32}$). These values point towards semi-elliptical profiles, sharply decreasing from the central peak with almost absent wings. This feature, already qualitatively noted from the column density maps (see Fig.~\ref{fig:OMC3showcase}, central panel), is a direct consequence of the density-selective nature of N$_2$H$^+$ as a tracer and a first argument supporting the Gaussian function as preferred fitting profile.

\begin{figure}
    \centering
    \includegraphics[width=\linewidth]{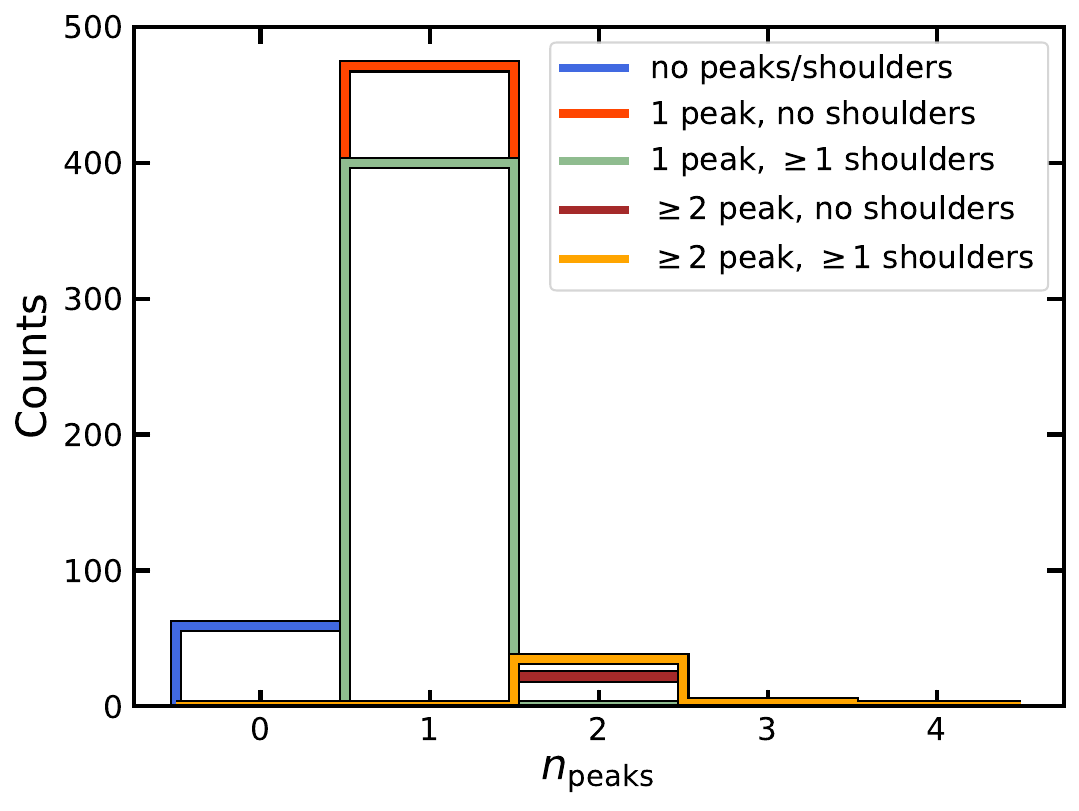}
    \caption{Histogram comprising the 991 profiles fitted by FilChap. Single-peaked ($n_\mathrm{peaks}$) profiles are the majority, with or without shoulders (green and red, respectively). Profiles with multiple significant peaks ($n_\mathrm{peaks}\geq2$, brown and orange) are a minority. The same applies for profiles below the significance criterion, but still with a convergent fit (blue).}
    \label{fig:npeaks}
\end{figure}

The next step in this quantitative description is to assess how representative is the fit for the profile. For this reason, we explore the number of peaks and shoulders identified by FilChap across our sample in Fig.~\ref{fig:npeaks}. Almost $\sim90\%$ of the selected cuts has only one significant (i.e. N$(\mathrm{H_2}) > 5\times\sigma$, see above) maximum or peak within the radially sampled distance. The number of cuts with mutiple peaks is instead only a minority ($\sim6\%$), as well as for those noisy cuts where no significant peak was identified, but a convergent fit was still possible (\textit{no peaks/shoulders} in Fig.~\ref{fig:npeaks}; $\sim5\%$). These results highlight how, despite the presence of a single or multiple shoulders, the majority of our radial cuts is well represented by a single, prominent structure.

\begin{figure}
    \centering
    \includegraphics[width=\linewidth]{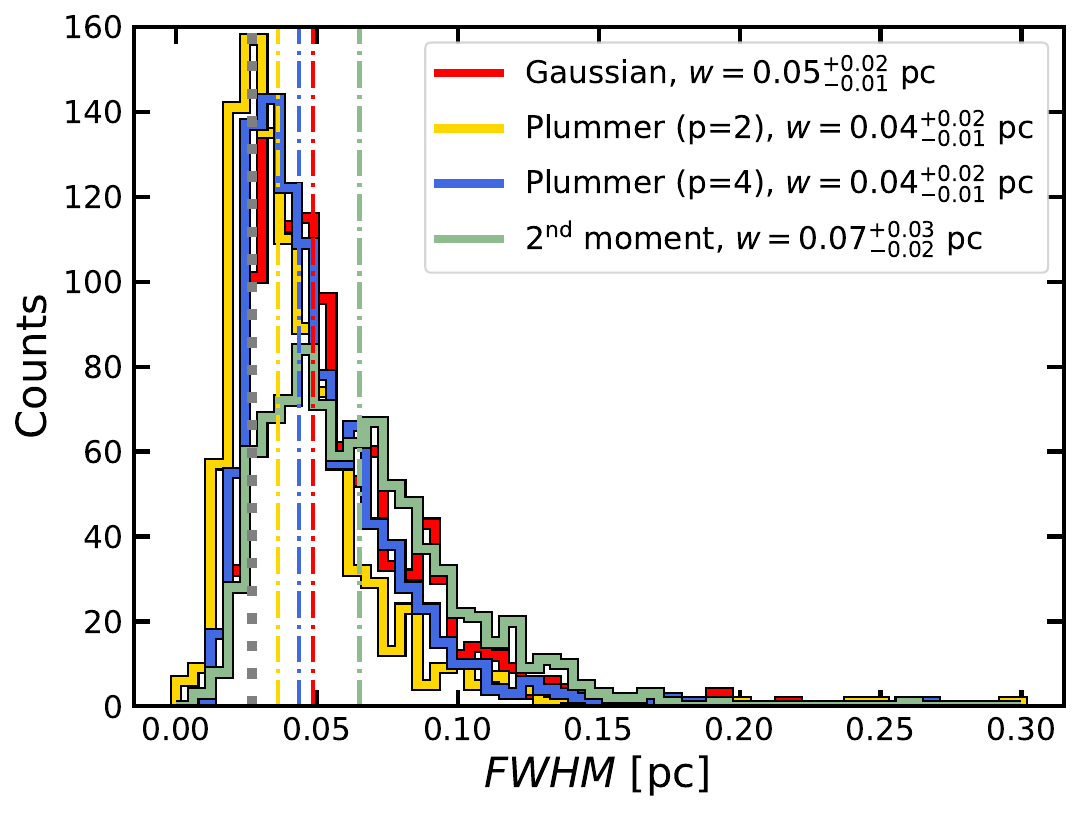}
    \caption{Distributions of widths for the individual cuts in our survey. These distributions are colour-coded based on the different estimates used to determine the width, Gaussian (red), Plummer $p=2$ (yellow), Plummer $p=4$ (blue), and $2^\mathrm{nd}$ moment of the distribution (green). The same colour-coding is applied for the median values of each distribution (vertical lines).}
    \label{fig:difffit}
\end{figure}

\begin{figure}
    \centering
    \includegraphics[width=\linewidth]{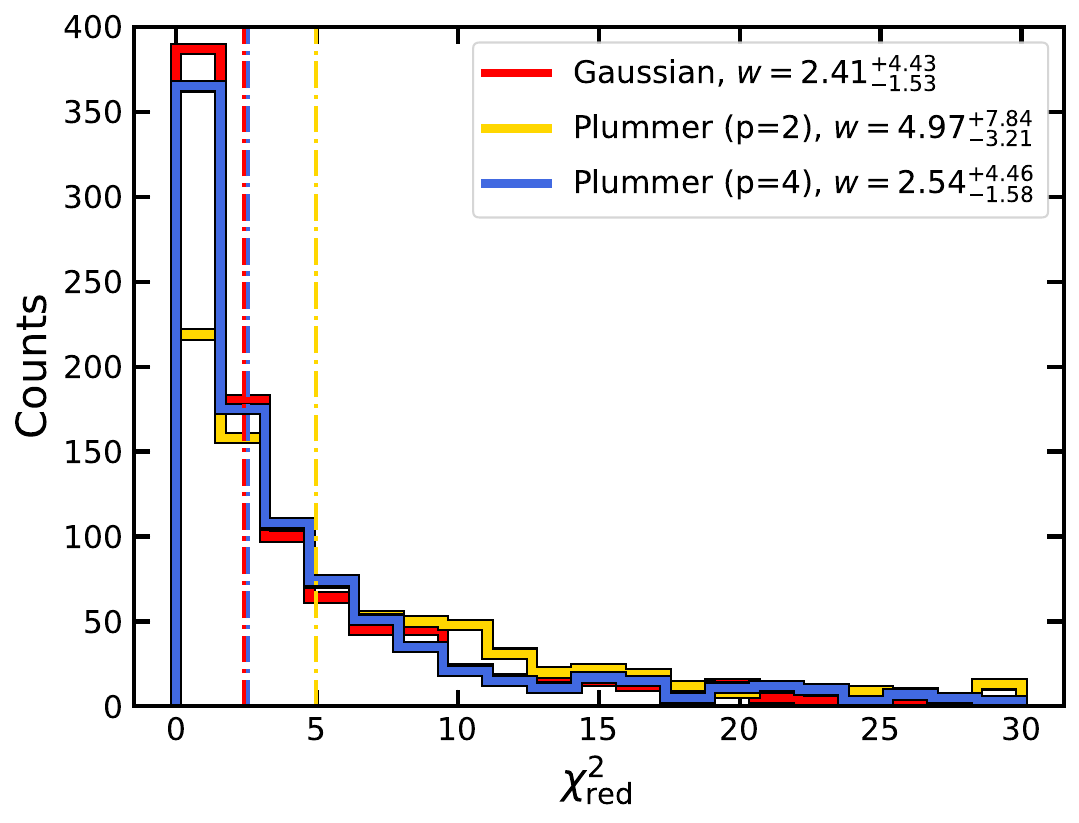}
    \caption{$\chi^2$ of the three fitting functions applied to our individual cuts, Gaussian (red), Plummer $p=2$ (yellow), and Plummer $p=4$ (blue). The median values for each distribution (vertical lines) bear the same colour-coding.}
    \label{fig:chi2}
\end{figure}

To describe the physical properties of these single-peaked structures, in particular their width and peak column density, we need a fit of the profile. FilChap provides two fitting functions to carry out this operation, the Gaussian and the Plummer. These are the two typical filament radial profile shapes determined in the past both in theory and observations (see Sect.~\ref{sec:theorobs}). We applied both to our radial cuts in FilChap, although we limited the number of free parameters for the Plummer function by fixing the $p$ exponent. By performing the fit for $p=2$ and $p=4$, we effectively reduce the degeneracy of the function on a single radial cut, whose sampling could be shallow, and thus ease the fit convergence.

The results from the three fits (Gaussian, Plummer ($p=2$), Plummer ($p=4$)) are compared in Fig.~\ref{fig:difffit} and Fig.~\ref{fig:chi2}. Figure~\ref{fig:difffit} shows the distribution for the widths obtained with the the three fitting functions. These distributions are all similar and extended for more than an order of magnitude ($\sim0.01-0.3$~pc). Although extended in range of values, the median widths of the three fits agree well within $\sim0.04-0.05$~pc with the majority of cuts below 0.1~pc, typical width for parsec-scale filaments. No fitting function appears as preferred based on these distributions, however, we have also to consider how many of these fits are reliable at our resolution. Considering 0.027~pc as resolution limit (dotted grey line; $3\times\theta_\mathrm{beam}$ at 414~pc), the Gaussian fit recovers only $\sim7\%$ of unresolved fits, while this number grows to $\sim15\%$ and $\sim30\%$ for the Plummer fits ($p=4$ and $p=2$, respectively).

Figure~\ref{fig:chi2} shows the reduced $\chi^2$ calculated for each cut as additional quality assessment for the fit. All three distributions peak within the first bin; however, the Plummer ($p=2$) shows a significant tail of values $\chi^2>5$ compared to the other two. This feature is reflected in the median values, for which the Gaussian and the Plummer ($p=4$) agree well with $\sim2.4$ and $\sim2.5$, respectively, while the Plummer ($p=2$) shows a median value doubled compared to the latter two (i.e. $\sim5$).

Based on the previous tests, the Plummer ($p=2$) function appears as the worst choice to fit the radial profiles of the Orion fibers. The other two fitting functions, namely Gaussian and Plummer ($p=4$), show extremely similar performances, both in terms of percentage of resolved widths recovered and accuracy in reproducing the profiles. The goodness of these two specific fits constitutes further proof of the sharpness of the profiles. Since we artificially enforced the $p$ exponent in the Plummer fit to reduce its degeneracy, we opt for the Gaussian function as fitting function of choice to discuss the results in the main text.

As additional test on the goodness of our fits, we retrieved the $\mathrm{2^{nd}}$ moment of the radial profiles (i.e. their variance) from FilChap. The variance informs us on the spread of the column density radial cut, beyond the width determined from the fit of its central peak. Figure~\ref{fig:difffit} shows the distribution of variances for the cuts of the fibers in Orion. While the values are more spread compared to the fitting results, the variances of the cuts are still for the vast majority below 0.1~pc ($\sim80\%$), exhibiting a median value of $\sim0.07$~pc, and consistent, within errors, with all the fit median values. The $\mathrm{2^{nd}}$ moment of our radial profiles, on the one hand, provides a more detailed information on the overall column density width in each cut, but, on the other hand, loses the information on the peak column density of the central condensation. Since one of our aims is to explore the correlation between $FWHM$ and $N_0$ and given the close agreement of the median values, we proceed in discussing the Gaussian fit results throughout the text.

\section{Additional maps and tables}\label{sec:mapsandtab}

\begin{figure*}[tbp]
\centering
\includegraphics[width=0.95\linewidth]{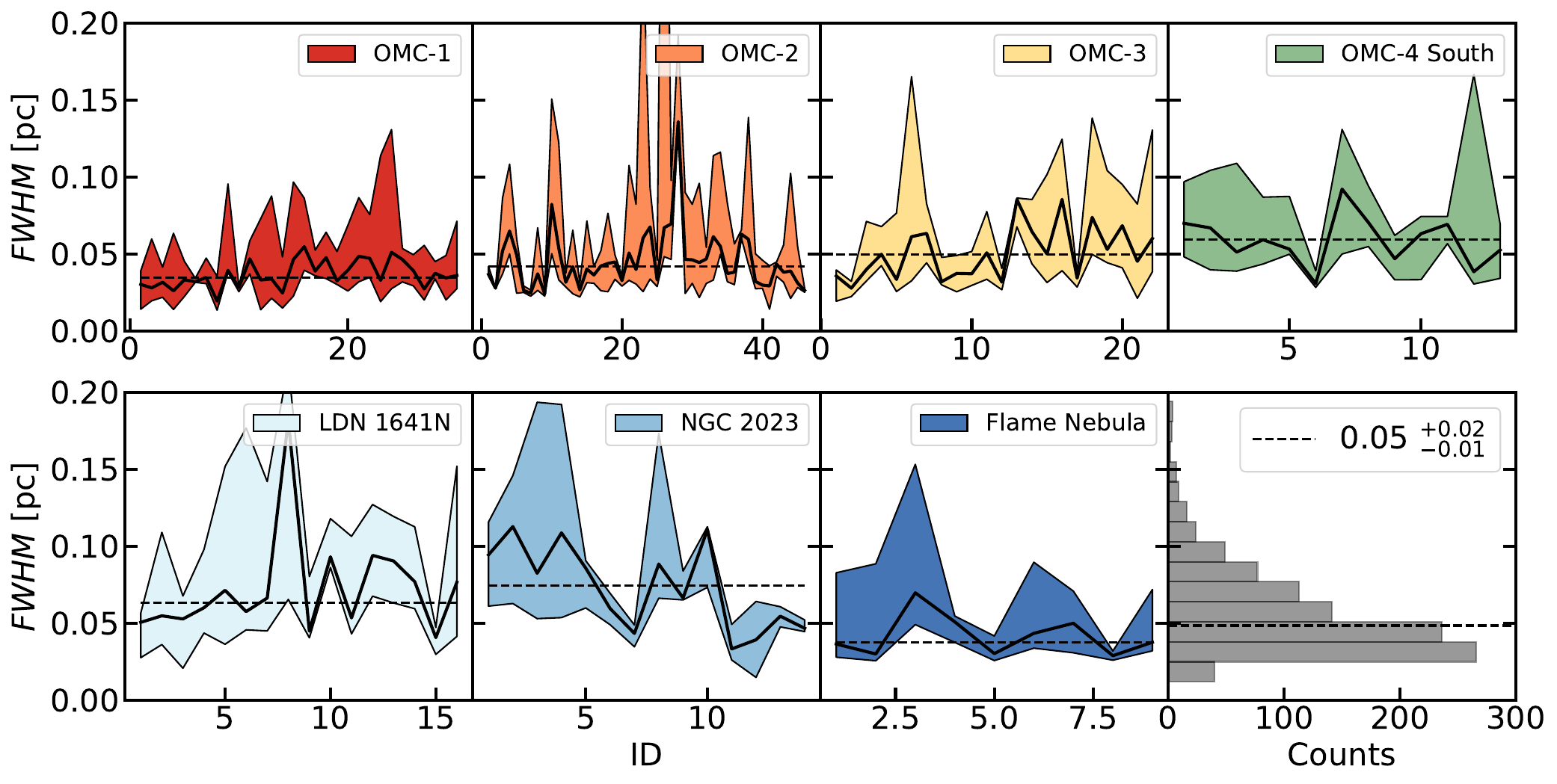}
\caption{We report here the variation in the width per fiber in each region composing the EMERGE Early ALMA Survey. The fibers are numbered based on their identification number (ID) and we display the range within minimum and maximum width per fiber with colour-coded areas. The solid black lines represent the median width per fiber, while the dashed lines the median width per region (e.g. circles displayed in Fig.~\ref{fig:filcompar}). The lower rightmost panel shows the distribution of all the 991 cuts fitted in our sample (see also Fig.\ref{fig:cumulativerad}, left panel).}
\label{fig:allcutsvar}
\end{figure*}

In Sect.~\ref{sec:fiberFWHM}, we argued towards the extreme variety of physical conditions, and, specifically, fiber widths when considering the distribution of all 991 radial cuts fitted by FilChap. However, while Fig.~\ref{fig:cumulativerad} provides a description of the widths variation within a region and in the sample as a whole, it also loses track of the intrinsic width variation per fiber. 
Figure \ref{fig:allcutsvar} shows the widths per cut, assigned to the corresponding fiber within each region. Each fiber is labelled with a corresponding ID during the identification process carried out by HiFIVe \citep[see][]{hacar18} and its narrowest and broadest cut are the extremes of the colour-coded area. The solid black line represents the median widths per fiber, discussed in Sect.~\ref{sec:fiberFWHM}, while the dashed black line the median width per region (see Table \ref{tab:gen_prop}). Finally, the lower, rightmost panel shows the distribution for the widths of all the cuts, along with the median value (see Table \ref{tab:gen_prop}).

As was discussed throughout the main text, our analysis is able to harness the complex morphology, the high dynamic range and spectral richness of the dense gas presented in Paper III (see also Fig.~\ref{fig:OMC3showcase}). The fitting of the fiber radial profiles per cut allows us to explore in depth the local properties of our fibers, with prime focus on the fiber widths. Across the survey, each region shows a unique distribution of such widths: regions such as OMC-1, OMC-4 South, and the Flame Nebula show constrained values, rarely exceeding $\sim0.1$~pc; regions such as OMC-2, OMC-3, LDN~1641N and NGC~2023, on the other hand, show more diverse distributions with several fibers showing almost an order of magnitude variation in their cuts ($\sim0.02-0.2$~pc). This large variety in width distributions, even within a single fiber, reinforces our claim around the absence of a typical width for fibers and, instead, its dependency with the environmental conditions of the host region.

\end{appendix}

\end{document}